**Title**: Deciphering the internal conversion and triplet formation in thymine via time-resolved multi-center X-ray photoelectron spectroscopy

Xiaojun Wang,[1*] Woojin Park,[2] Dennis Mayer,[1] Fabiano Lever,[1] Giovanni Cirmi,[1] Michele Di Fraia,[3] Benjamin Erk,[1] Marion Kuhlmann,[1] Zane Phelps,[4] Oksana Plekan,[5] Federico Pressacco,[1] Sebastian Schulz,[1] Cheol Ho Choi,[2] Miquel Huix-Rotllant,[6*] Markus Gühr[1,7*]

[1]Deutsches Elektronen-Synchrotron DESY, 22603 Hamburg, Germany

[2]Department of Chemistry, Kyungpook National University, Daegu 41566, South Korea

[3]CNR-Istituto Officina dei Materiali (IOM), Basovizza, Trieste 34149, Italy

[4]James R. Macdonald Laboratory, Physics Department, Kansas State University, Manhattan, 66506, KS, USA

[5]Elettra-Sincrotrone Trieste S.C.p.A., Basovizza 34149 Trieste, Italy

[6]Aix-Marseille Université, CNRS, ICR, Marseille 13397, France

[7]Institute of Physical Chemistry, Universität Hamburg 20148 Hamburg, Germany

*Corresponding author: xiaojun.wang@desy.de miquel.huix-rotllant@cnrs.fr markus.guehr@desy.de

## Abstract

The photostability of DNA nucleobases relies on ultrafast nonradiative relaxation following ultraviolet excitation. In thymine, the competition between internal conversion (IC) and intersystem crossing (ISC) determines the balance between ultrafast energy dissipation and potentially long-lived photochemistry. However, the precise molecular mechanisms linking early singlet-state electronic and structural evolution to the formation of long-lived triplet dark states remain unresolved. Here, we investigate the gas-phase photodynamics of thymine using time-resolved multi-center X-ray photoelectron spectroscopy (tr-mc-XPS) supported by state-of-the-art *ab initio* quantum calculations. By simultaneously probing multiple carbon 1s core levels, we track site-specific electronic and nuclear dynamics. We discover that the long-lived dark state is not initially a pure triplet; rather, it exists as a mixture of singlet and triplet states for the first 240 ps before transitioning fully to the $^3\pi\pi^*$ state. On the femtosecond timescale, we resolve atom-specific vibrational coherence at ~730 cm$^{-1}$ (a ring-breathing mode with a period of 46 fs) that survives the initial $^1\pi\pi^*$ to $^1n\pi^*$ IC. Furthermore, we identify a critical dual role for the methyl group ($C9H_3$). During singlet relaxation, the methyl group acts as an inertia spectator, and its mass hinders direct internal conversion to the ground state. Upon $^3\pi\pi^*$ state formation, however, state-dependent hyperconjugation couples the methyl group to the pyrimidine ring, turning it into a highly sensitive electronic reporter of ISC. These findings establish tr-mc-XPS as a powerful approach for disentangling the complex, multi-timescale photodynamics governing nucleobase stability.

## Introduction

Nucleobases absorb strongly in the ultraviolet (UV) spectral region, yet they exhibit remarkable photostability through ultrafast nonradiative relaxation pathways.[1–4] This ability to efficiently dissipate UV excitation energy is central to the photochemical stability of genetic material and has long motivated the study of nucleobase photophysics. The balance between efficient energy dissipation and photochemical reactivity is particularly important for thymine, whose UV excitation can initiate DNA photodamage. The most abundant UV-induced DNA lesion, cyclobutane pyrimidine dimer (CPD),[5] can form on ultrafast timescales immediately following thymine UV excitation to a bright singlet state.[6] Triplet-mediated pathways provide an additional route to CPD formation on longer timescales through triplet-triplet energy transfer.[7] Understanding how thymine distributes its electronic excitation among singlet and triplet states is therefore essential for establishing the molecular mechanisms underlying both photostability and photodamage.

The photophysics of thymine involves a cascade of coupled electron and nuclear dynamics, including internal conversion (IC) and intersystem crossing (ISC). Thymine absorbs broadly across the UVB and UVC wavelength range, with an absorption peak ~ 250 nm corresponding to excitation to the $^1\pi\pi^*$ state.[8] In DNA, the electronic structure and relaxation of thymine is complex due to interaction with neighboring nucleobases and the other constituents of the DNA structure.[3] In liquids, a direct, sub 100 fs path to the ground state ($S_0$) collects about 80% of the excited population, with the remainder undergoing IC into the $^1n\pi^*$ state.[9,10] Gas phase experiments take an important role in elucidating the intrinsic photophysics of thymine, free from environmental perturbations, while enabling direct comparison to high-level *ab initio* calculations.[11–13] In the isolated case, direct IC decay to the ground state is not observed; instead, all population relaxes into the $^1n\pi^*$ state (Figure 1a), based on estimates in previous studies.[10,14,15] Miura et al. attribute the absence of this direct ground-state channel to the central role of the pyramidalization at C5, which is slower for thymine than for uracil due to the inertia of the methyl group, leading to negligible ground state relaxation in thymine compared to the $^1\pi\pi^*\rightarrow{}^1n\pi^*$ IC.[10] Besides fast IC dynamics, much longer timescales beyond 100 ps are observed in optical pump-probe spectroscopy in gas phase.[10,15–19] These so-called ‘dark states’ have been frequently associated with triplet states based solely on its lifetime.[16,18]

The triplet assignment has been a matter of discussion by a variety of theoretical and experimental approaches. *Ab initio* calculations by Serrano-Andres and coworkers pointed out a $^3\pi\pi^*$ triplet state intersystem crossing.[20] Ligare et al. provided one of the first experimental evidence by measuring N–H stretching frequencies of the dark state, which proved compatible with the $^3\pi\pi^*$ assignment.[19] Further support came from time-resolved (tr-) UV photoelectron spectroscopy; here, combining UV measurements with *ab initio* quantum dynamics simulations of efficient ISC further confirmed the triplet long-lived nature.[21,22] Although UV photoelectron spectroscopy could potentially miss populations that have relaxed to low-lying electronic states due to lower photon energy for ionization, time-resolved extreme ultraviolet (tr-EUV) photoelectron measurements revealed very similar long-lasting dark states.[10,15] Based on a simulation of the $^3\pi\pi^*$ EUV-photoelectron spectrum, this state was once more attributed to a triplet state.[15] Despite this accumulating consensus, the molecular mechanism connecting the early singlet-state dynamics to the formation and subsequent evolution of the long-lived triplet population remains unclear.

Here, we combine time-resolved core level X-ray photoelectron spectroscopy (XPS) at multiple carbon atoms with state-of-the-art *ab initio* quantum calculations. This route leads to clear evidence to the triplet interpretation of the long-lasting dark state and the molecular mechanisms leading to this state. We furthermore show that the long-lived ‘dark state’ is not described by a unique triplet signature, but instead exhibits two distinct temporal regimes in the XPS response. In addition, we resolve atom-specific vibrational coherences within the $^1n\pi^*$ state. The dominant oscillatory signal, consistent with previously reported ring-breathing mode,[23,24] captures real-time structural dynamics along the $^1n\pi^*$ potential energy surface. Finally, we reveal a dual role of the methyl group: its inertia suppresses the nuclear motion required for direct internal conversion to the ground-state, whereas its electronic coupling to the π system of the heterocycle (hyperconjugation) becomes pronounced upon triplet formation.

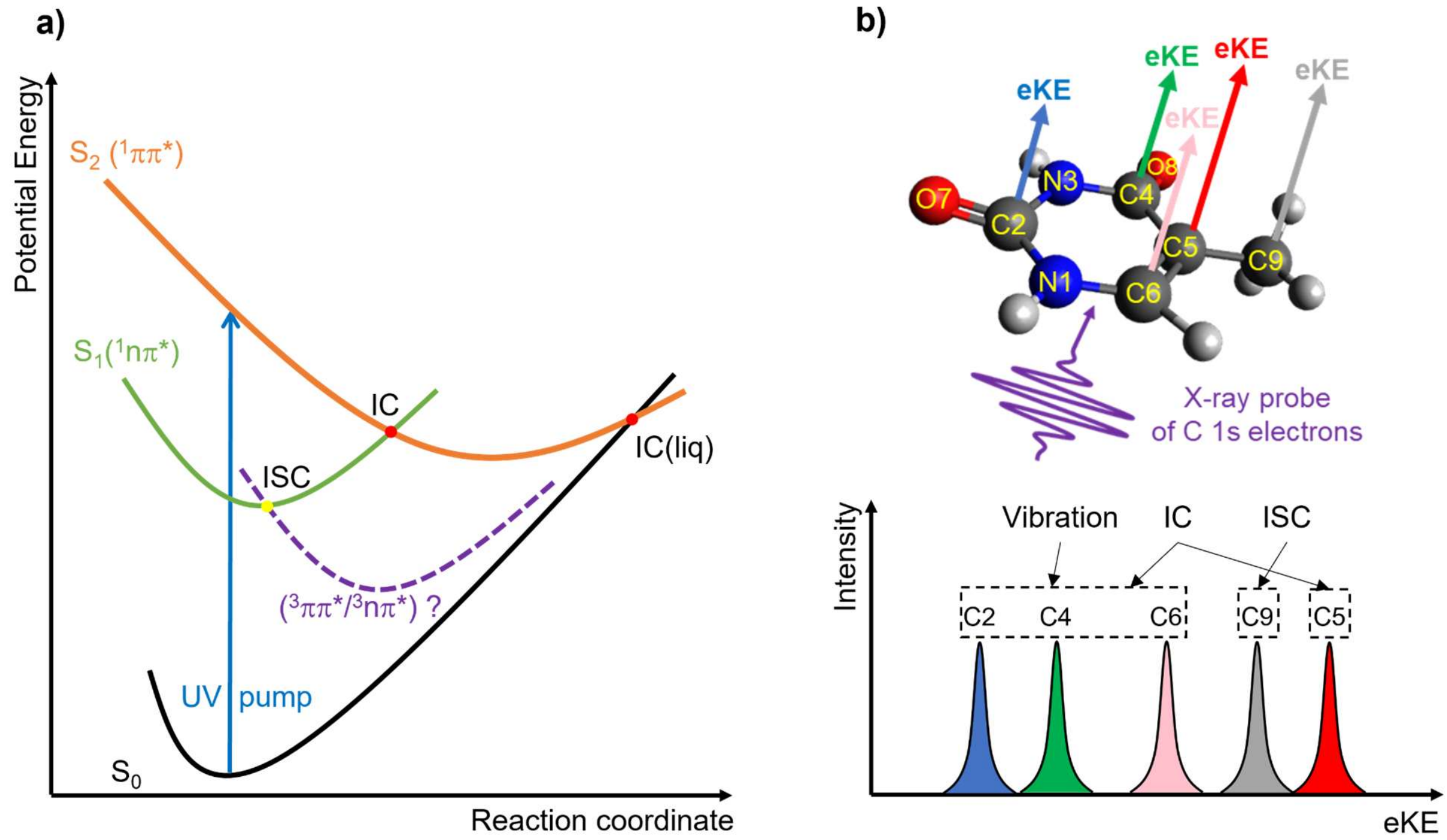

**Figure 1.** Scheme of the thymine electronic relaxation and the tr-mc-XPS probing mechanism. (a) Non-radiative relaxation pathways of thymine. In the gas phase, UV absorption promotes thymine into the $S_2$ ($^1\pi\pi^*$) state, which relaxes to the $S_1$ ($^1n\pi^*$) state (IC; red dot) and subsequently undergoes intersystem crossing (ISC; yellow dot) to long-lived triplet states. In the liquid phase, however, approximately 80% of excited thymine molecules bypass this path and return directly to the ground state (IC (liq); red dot). (b) The tr-mc-XPS probing scheme. A UV pump excites thymine to the $^1\pi\pi^*$ state and an X-ray probe ionizes the C 1s core levels. The top model illustrates photoelectrons emitted from the ring (C2, C4, C5, C6) and methyl ($C9H_3$) carbons. The bottom scheme displays the corresponding conceptual photoelectron spectrum. Excited-state valence electron redistribution changes C 1s screening, shifting the electron kinetic energy (eKE) of these distinct peaks. As annotated, these shifts encode specific dynamics: the C9 shift indicates ISC; the combined shifts of C2, C4, C5, and C6 indicate IC; and the combined shifts of C2, C4, and C6 indicate an excited-state vibrational coherence.

In this paper, we also show how tr-XPS can be extended to multiple atomic centers, thereby delivering even richer information on the molecular dynamics. Static XPS is element sensitive, has an advantage of high local sensitivity due to the core-hole localization, and allows to distinguish sites via a small eV-scale chemical shift.[25] The tr-XPS technique is established in gas phase molecules and can give new insight on electronic dynamics from a local-electronic perspective.[26–31] Shifts in the tr-XPS can be interpreted by changing local charges due to optical excitation, giving rise to the term 'excited state chemical shift' (ESCS) effect.[28,31] Here we apply tr-XPS on all five carbon 1s levels of thymine, introducing an approach we name time-resolved multi-center XPS (tr-mc-XPS). Figure 1b provides an overview of this method probing carbon atoms C2, C4, C5, C6, and C9, which encode distinct dynamical information about the molecule. Specifically, the kinetic energy shift of the C9 core level provides a direct probe of intersystem crossing. And atoms C2, C4, C5, C6 act as reporters on the $^1\pi\pi^* \rightarrow {}^1n\pi^*$ IC and also provide a view of the molecular vibrational dynamics along the IC and subsequent $^1n\pi^*$ evolution. This multi-center capability establishes a new molecular fingerprint in XPS, where electronic and nuclear dynamics are mapped across spatially separated sites.

Combining these multi-center experimental observables with quantum-dynamical simulations ultimately allows us to disentangle the coupled electronic and nuclear processes governing thymine photochemistry.

## Method

### Experimental method

The time-resolved UV pump–soft X-ray probe photoelectron spectroscopy of gas-phase thymine was conducted at the FL24 beamline of the FLASH free electron laser (FEL) at DESY,[32] using the URSA-PQ end station. A detailed documentation of the apparatus and experiment can be found elsewhere.[33] The experimental scheme is summarized in section S1 of the supporting information (SI). In short, the apparatus includes a 1.7 m long time-of-flight magnetic bottle electron spectrometer (MBES) and a capillary oven[34] for sample delivery. Thymine molecules were evaporated from the oven, heated up to 180° C and guided to the interaction region through an aluminum capillary. The molecular density in the interaction region was estimated to be $1.7 \times 10^{13}$ $cm^{-3}$.[34]

The X-rays, with a mean photon energy of 337.6 eV and a bandwidth of approximately 1.9 eV (full width at half maximum, FWHM), were used to probe the dynamics by carbon 1s core-level ionization. The soft X-ray probe pulses were generated via self-amplified spontaneous emission (SASE), with an effective repetition rate of 1 kHz. Each pulse had a few μJ pulse energy and a duration of ~50 fs and was focused to a spot size of ~50 μm (FWHM). Thymine molecules were excited to the bright $S_2$ ($^1\pi\pi^*$) state with an ~80 fs UV pump pulse at 263 nm (see Figure 1a). The pump pulse, with an estimated upper limit for the pulse energy of 1.5 μJ, was focused to a spot size of ~100 μm (FWHM), ensuring that the molecules were not overpumped and the experiment took place in linear absorption regime.[14] Both the X-ray and UV pulses were linearly polarized along the axis of the MBES.

The temporal delays between UV pulse and X-ray were sampled on a non-uniform grid with fine steps near temporal overlap and coarser spacing at long delays. Around time zero, delays were scanned with 20 fs steps, providing high resolution of ultrafast dynamics. This was followed by increasing step sizes from sub-ps to several ps. At longer delays, the step size grew quasi-logarithmically, reaching tens to hundreds of ps, with the maximum delay extending to 1 ns. This delay scheme enabled us to capture the molecular dynamics of photoexcited gas-phase thymine over a wide range of time delays.

Photoelectron spectra as a function of the delays were recorded by the MBES. To improve the energy resolution of the spectrometer, the speed of the photoelectrons (around 45 eV) was reduced to 22 eV by an -23 V retardation voltage. The UV pump laser was off on every second X-ray pulse, facilitating fast difference spectra accumulation during the live data acquisition. Time-resolved difference spectra of the excited molecules were then obtained by subtracting the UV-off spectra from the UV-on spectra at various delays. We applied a temporal jitter correction to the X-ray pulses relative to delays set during data acquisition, and subsequently rebinned the delays to achieve both equal statistics per delay to improve the temporal resolution. The temporal overlap and instrument response function have been determined by analyzing the integrated difference signal (see section S2 of the SI for further details). The instrument response function (IRF) was estimated to have a standard deviation of (67 ± 5) fs, corresponding to a FWHM of (158 ± 12) fs.

### Computational details

All electronic-structure calculations were performed using OpenMolcas 24.06[35] with the ANO-RCC-VDZP basis set. Ground- and excited-state minima ($S_0$, $S_1$, $T_2$, and $T_1$), as well as relevant electronic state crossings, were optimized at the 10-state-averaged extended multi-state complete active space second-order perturbation theory (10SA-XMS-CASPT2) level. The active space used for geometry optimizations and valence excited-state calculations comprised 14 electrons in 9 orbitals, including five π, two n, and two π* orbitals as shown in Figure S7 in section S6 of the SI. Harmonic frequency calculations were performed to confirm the optimized minima. Potential energy curves were evaluated at the 12SA-XMS-CASPT2(14e,9o)/ANO-RCC-VDZP level, which was found to best reproduce the experimental electron energy loss spectroscopy (EELS) excitation energies of the two lowest singlet and the two lowest triplet states in the gas phase.[36]

Carbon 1s X-ray photoelectron spectra at the optimized $S_0$, $S_1$, $T_2$, and $T_1$ minimal energy geometries were computed using extended multi-state restricted active space second-order perturbation theory (XMS-RASPT2). Different active spaces from those for the valence excitations were used for the XPS calculations, tailored to the target state manifold at each geometry: (12e, 9o) (one n, five π, three π*) at the $S_0$ and $S_1$ minima, and (14e,10o) (two n, five π, three π*) at the $T_1$ and $T_2$ minima. These active spaces were shared among all neutral and core-ionized states calculated at the respective geometries. Note that the carbon 1s core orbitals in the RAS1 space were not counted here. The RASSCF orbitals were separately optimized in state-averaged calculations including 10 singlet, 10 triplet, and 15 core-ionized doublet and quartet states. For all perturbative calculations, an imaginary shift of 0.35 Hartree was applied to avoid intruder state problems, and the IPEA shift was set to zero.[37]

The simulated XPS spectra were constructed by convolving the computed vertical ionization energies with the corresponding Dyson orbital strengths using a Gaussian broadening of 0.35 eV. Dyson orbitals were evaluated using XMS-RASPT2 energies and perturbatively mixed RASSCF transition densities. The active-space and basis-set dependence of the calculated XPS spectra were carefully benchmarked in SI, and the XPS-specific flavor was selected to reproduce the experimental ground-state carbon 1s spectrum. Details of the active-space selection, spectral shifts, Dyson orbital analysis, Cartesian XYZ coordinates, and XPS benchmarking are provided in the SI.

## Results

### Experimental results

#### Static spectrum

Figure 2a shows the experimental carbon 1s photoelectron spectrum of gas-phase thymine in the ground state. The black line represents the spectrum obtained in the present experiment using X-ray radiation with a mean photon energy of 337.6 eV. We shift the experimental spectrum to fit the broadened literature spectrum, obtained by Plekan et al. at a high-resolution synchrotron source.[38] Based on their experimental data and theoretical calculations, the carbon 1s photoelectron peaks (dashed lines) appear at 42.4 eV for C2, 43.4 eV for C4, 45.3 eV for C6, and 46.6 eV for both C5 and C9, as numbered in Figure 1b. The 1.9 eV bandwidth of the X-ray pulse in the current experiment prevents the resolution of individual photoelectron peaks for each carbon atom, resulting in two broad peaks in our spectrum.

**Difference spectrum**

Figure 2b shows the experimental tr-XPS C 1s difference spectrum of photoexcited thymine. The measurements cover time delays from a few tens of femtoseconds up to one nanosecond after photoexcitation, nearly six orders of magnitude, thus capturing the relaxation from the initial excited state to long-lived triplet state. The high repetition rate of the FLASH FEL enabled us to achieve an excellent signal-to-noise ratio in the difference spectrum within only two hours. In contrast to the unstructured static spectrum, the broad bandwidth of FEL does not hinder the identification of rich features in the difference spectrum. This allows the attribution of distinct intermediate electronic states. Four temporal regions (R1–R4) with distinct spectral features can be clearly identified in the difference spectra, as separated by horizontal dashed lines in Figure 2b. To emphasize the spectral variations characteristic of each temporal regime, the averaged spectra corresponding to selected delays within each region are displayed in Figure 2c.

Region R1, occurring within the first 250 fs, is characterized by a bleach centered at 45.5 eV and a UV-induced positive tail extending into the higher eKE region from 48.5 eV to 52.5 eV. In addition, the bleach and the UV-induced positive feature span the eKE regions 41–44 eV and 46–48.5 eV, respectively. A similar tail has been reported in previous time-resolved valence PES experiments.[10,14,24] The electronic state associated with this high eKE tail feature has been identified as the $S_2$ ($^1\pi\pi^*$) state, which is populated immediately after photoexcitation. Thymine subsequently relaxes out of the Franck–Condon (FC) region within ~200 fs via an ultrafast internal conversion from the $S_2$ ($^1\pi\pi^*$) to the $S_1$ ($^1n\pi^*$) state.

In R2, the bleach at 45.5 eV disappears, while the bleach (41–44 eV) and the UV-induced positive features (46–48.5 eV) observed in R1 persist up to 7 ps. Based on the temporal scale of R2 that is consistent with the reported lifetime of the dark $S_1$ ($^1n\pi^*$) state,[10,16,18,21] the spectral features in this region are attributed to the $S_1$ ($^1n\pi^*$) state of thymine.

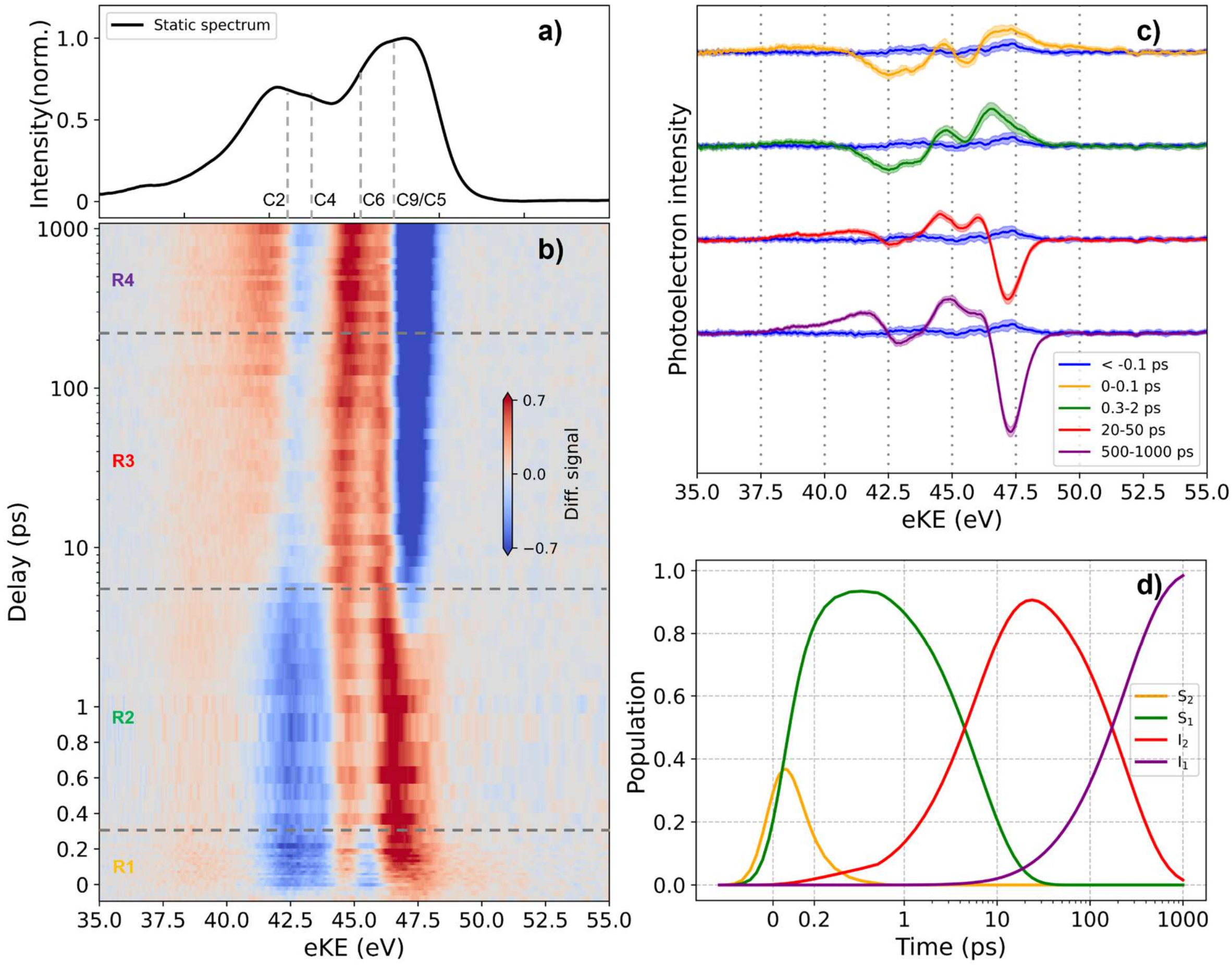

**Figure 2.** (a) Experimental carbon 1s photoelectron spectrum of gaseous thymine in its ground state. The previously reported data[38] (dashed lines) indicate the photoelectron peaks of individual carbon atoms. (b) Difference spectra as a function of pump–probe delay, spanning from tens of femtoseconds to several hundred picoseconds, revealing the full relaxation dynamics. The spectra are divided into four regions (R1–R4) based on distinct spectral features. (c) Averaged difference spectra at selected delays for each region shown in (b), highlighting characteristic spectral signatures; uncertainties are estimated using a bootstrap method. (d) Time evolution of population extracted from the difference spectra in (b) using global analysis with a four-state cascade kinetic model. The extracted lifetimes are (92 ± 18) fs for $S_2$, (6.3 ± 0.2) ps for $S_1$, and (239 ± 29) ps for $I_2$, where the uncertainties represent one standard deviation; the $I_1$ lifetime exceeds 1 ns and lies beyond the experimental time window. The uncertainties of the lifetime were determined using bootstrap analysis combined with the global fitting procedure.

In contrast to R1 and R2, regions R3 and R4 show a pronounced change near 47.0 eV. Here the positive feature seen in R1 and R2 becomes negative; this negative feature persists up to the end of the measured delay range (1 ns). R4 is further characterized by an enhancement of the positive signal in the range of 40.0–42.5 eV and a negative signal from 46.5 to 48.5 eV. Comparing the average spectra in R3 and R4 (Figure 2c, red and purple lines, respectively) reveals that the two positive features observed in R3 within the 44.0–46.5 eV range merge into a single broad peak in region R4, clearly distinguishing it from R3. R3 persists for nearly 200 ps, whereas R4 extends beyond the temporal delay range (1 ns) recorded here. Previous studies have generally attributed such long-lived states to triplet states.[10,14,15,18,21]

To quantitatively extract the lifetimes of electronic states, we employ a global analysis, which decouples the time-resolved difference spectra into the state-associated spectrum and corresponding population dynamics.[39,40] We adopt a kinetic model in which four sequentially populated states participate: $S_2$, $S_1$ states and two more states $I_2$ and $I_1$. The latter two, we will attribute to a mixed singlet-triplet state and a pure triplet state in the discussion. The lifetimes of all states, together with the instrument response function, are treated as fitting parameters to minimize the residuals between the experimental difference spectrum and the optimized fitted spectrum. The validity of the four-state kinetic model is confirmed by a significant reduction in the systematic residuals compared to a three-state model (see section S3 of the SI for details on kinetic model selection).

Figure 2d presents the resulting population dynamics of each state. The extracted lifetimes are (92 ± 18) fs and (6.3 ± 0.2) ps for $S_2$ and $S_1$ states, respectively. The $I_2$ exhibits a lifetime of (239 ± 29) ps, while $I_1$ persists beyond 1 ns, exceeding the experimental time window. The instrument response function is characterized by a standard deviation of $\sigma$ = (72 ± 5) fs, consistent with the value of (67 ± 5) fs determined from the integrated intensity as described in the experiment section. Uncertainties were estimated using a bootstrap approach combined with global analysis.

We note that the $S_2$ lifetime of (92 ± 18) fs is in agreement with the lifetime found by Wolf et al.[14] of (60± 30) fs but does not overlap within a standard deviation with the recent findings from Miura et al.[10] of (39±1) fs. The data from Miura et al. were taken by valence photoelectron spectroscopy, whereas the soft X-ray measurements might have a slightly different observation window on the $S_2$ population. In particular, the two techniques could differ in their sensitivity to the molecular distribution with respect to energy or internuclear separation in the $S_2$ state induced by the different pump wavelength (263 nm in the present experiment vs. 267 nm in Miura et al.) as well as in the ionization cross sections.

**Photoelectron intensity oscillations at early delays**

Figure 3a presents a magnified part of R1 in the difference spectrum shown in Figure 2b, where pronounced oscillations in the photoelectron intensity are observed across the entire eKE range. These oscillations persist within the first 200 fs after photoexcitation and are not visible at later delays most likely due to an increased delay step size. First, we concentrate on the eKE range of 41–44 eV, which corresponds to the bleach feature. Taking the absolute value of this range (visible in Figure 3b), the photoelectron intensity reaches successive minima at approximately 43, 82, 135, and 185 fs, as indicated by the red dashed lines in Figure 3a. Similar oscillatory behavior is observed in the 45–46 eV eKE region associated with the second bleach feature.

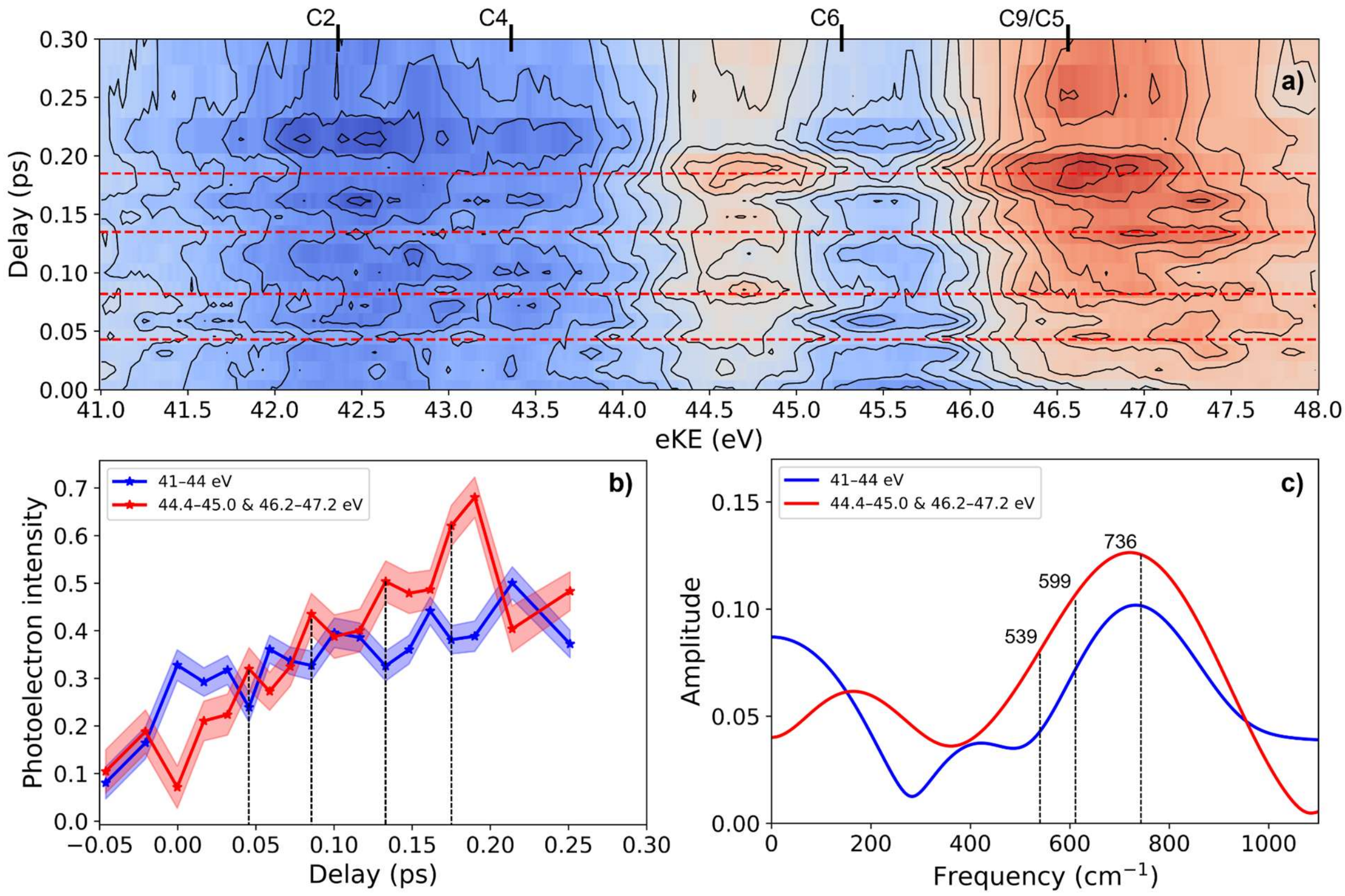

**Figure 3.** (a) The contour plot of the magnified region of the difference spectrum. Red dashed lines at approximately 43, 82, 135, and 185 fs mark the oscillation minima and maxima, corresponding to the ground-state bleach regions and the UV-induced increase regions, respectively, with a phase shift of π. Tick marks labeled C2, C4, C5, C6, and C9 indicate the C 1s photoelectron peaks of individual carbon atoms in unexcited thymine. (b) Photoelectron intensity integrated over the representative eKE regions (41–44 eV, 44.5–45 eV & 46.2–47.2 eV); black dashed lines highlight the out-of-phase oscillatory behavior observed in panel (a). (c) Frequency components of the photoelectron intensity oscillations in panel (b), obtained by Fast Fourier Transform (FFT) analysis over a window of the first 200 fs. The dominant frequencies of the representative regions are similar to one another and close to the previously reported ring-breathing mode frequency of ~750 cm$^{-1}$. The three calculated ground-state normal-mode frequencies (539, 599, and 736 cm$^{-1}$) are indicated by dashed lines. Peaks near these frequencies are prominent in the $S_1$ ($^1n\pi^*$) population and are addressed in the following sections.

In contrast, for the remaining eKE regions exhibiting UV-induced intensity increases, the photoelectron signal reaches maxima at the same delays, indicating a phase shift of π relative to the signal in the eKE ranges of 41-44 eV & 45-46 eV in Figure 3b. This out-of-phase behavior is more clearly illustrated in Figure 3b, which shows the photoelectron intensity integrated over the representative eKE regions as a function of pump-probe delay. The corresponding minima and maxima are clearly resolved, and the error bars are estimated using a bootstrap method.

To extract the oscillation frequencies, a Fast Fourier Transform (FFT) analysis was applied to the integrated photoelectron intensities over a window of the first 200 fs (Figure 3b). The resulting frequency spectra, shown in Figure 3c, exhibit a similar dominant component: ~730 cm$^{-1}$ for the 41–44 eV region and ~720 cm$^{-1}$ for the 44.5–45 eV and 46.2–47.2 eV regions. These frequencies are close to the previously reported ring-breathing mode frequency of ~750 cm$^{-1}$ (~44 fs in period) observed in the time-resolved photoelectron spectra of gaseous thymine.[24] The observation of this coherent motion in a site-sensitive core-level observable demonstrates that tr-mc-XPS can report not only on electronic-state populations but also on the real-time nuclear motion that accompanies their evolution.

## Theoretical results

### Relaxation pathway rationalized by potential energy curve

We present theoretical simulations of the molecule to support our subsequent discussion and enable conclusions beyond those drawn from experimental features alone. We start with a simulation on the relevant potential energy pathways governing molecular relaxation. Figure 4 depicts the potential energy curves of thymine computed at the 12SA-XMS-CASPT2(14e,9o)/ANO-RCC-VDZP level of theory along a relaxation pathway from the initially populated $S_2$ FC state through the conical intersection region to $S_1$ ($CI_{21}$), followed by ISC via the $S_1/T_2/T_1$ three state crossing region and subsequent relaxation to the $T_1$ minimum. In addition, there is a $T_1/S_0$ crossing region with a barrier of 1.33 eV relative to the $T_1$ minimum.

At the FC geometry, the bright $S_2$ state is characterized as a $^1\pi\pi^*$ state. Following excitation, the system evolves rapidly away from the FC region toward a $CI_{21}$ region, driven primarily by bond length alternation (BLA) motion along the O8=C4–C5=C6 framework.[41] As a result, the $S_2$–$S_1$ energy gap decreases rapidly, enabling ultrafast internal conversion to the dark $S_1$ state. After population transfer to $S_1$, the system relaxes further to the $S_1$ minimum, which is characterized by a $^1n\pi^*$ configuration with enhanced BLA distortion. The $CI_{10}$ ($^1n\pi^*/S_0$) lies at 5.75 eV (see Figure S17 in section S16 of the SI), which is approximately 1.6 eV higher in energy than the $S_1$ minimum (4.15 eV). This large energetic separation hinders the nonradiative $S_1 \rightarrow S_0$ internal conversion pathway, resulting in trapping of the excited-state population in the $S_1$ minimum ($S_1$ trapping).

Instead, in the vicinity of the planar $S_1$ minimum, the $T_2$ ($^3n\pi^*$) state lies energetically very close, suggesting a possible ISC pathway. However, direct $S_1$ ($^1n\pi^*$) $\rightarrow$ $T_2$ ($^3n\pi^*$) spin–orbit coupling (SOC) is expected be weak according to El-Sayed's rule.[42] Consistent with this expectation, the computed $S_1/T_2$ SOC at the $S_1$ minimum is only 7.4 $cm^{-1}$, indicating that efficient ISC requires vibronic enhancement through symmetry breaking nuclear motions that mix $n\pi^*$ and $\pi\pi^*$ characters. Normal-mode scans around the $S_1$ minimum revealed that, along such promoting coordinates, $S_1$, $T_2$ and the lower-lying $T_1$ states become simultaneously close in energy, although these regions are located at relatively high energies (see Figure. S16 in section S14 of the SI). We therefore optimized $S_1/T_2/T_1$ three-state crossing point to identify lower-energy ISC pathways. The optimized $S_1/T_2/T_1$ three-state crossing point lies only ca. 0.5 eV above the $S_1$ minimum, making it energetically accessible from the $S_1$ minimum. Notably, this crossing point retains a nearly planar structure, with enhanced BLA distortion as the main structural change, rather than the out-of-plane distortion normally expected for an ISC process. At this geometry, the computed $S_1/T_1$ SOC is 37.0 $cm^{-1}$, substantially larger than the $S_1/T_2$ SOC of 10.7 $cm^{-1}$. This suggests that the three-state crossing region can mediate efficient population transfer from $S_1$ to $T_1$, rather than enforcing a strictly sequential $S_1 \rightarrow T_2 \rightarrow T_1$ mechanism. After the population reaches the $T_1$ minimum, the 1.33 eV barrier to the $T_1/S_0$ crossing strongly suppresses ground-state recovery, leading to long-lived trapping on the $T_1$ state. This picture establishes a mechanistic connection between the early singlet-state dynamics and the subsequent triplet population: the same nuclear landscape that suppresses direct ground-state recovery leaves thymine trapped in an excited-state manifold from which intersystem crossing becomes accessible.

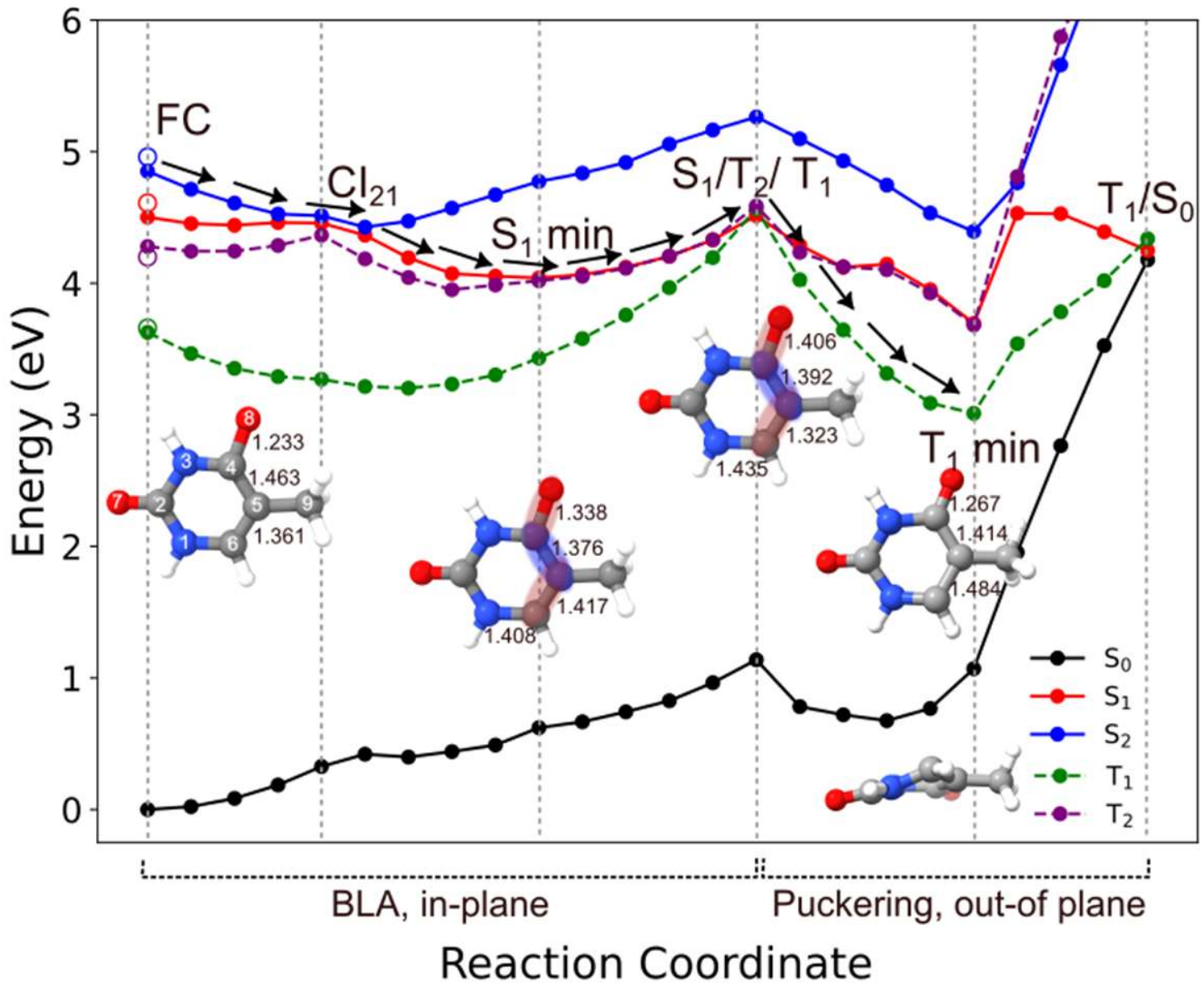

**Figure 4.** Potential energy curves of thymine along a representative excited-state relaxation pathway connecting the Franck–Condon (FC) region, $S_2/S_1$ conical intersection ($CI_{21}$), $S_1$ minimum, $S_1/T_2/T_1$ three state crossing, $T_1$ minimum, and $T_1/S_0$ crossing. The energies of $S_2$, $S_1$, $T_2$ and $T_1$ are plotted in blue, red, purple, and green, with singlet and triplet states represented by solid and dashed lines, respectively. Empty circles at the FC geometry denote vertical excitation energies from EELS.[36] Representative molecular structures and key structural distortions are shown along the pathway.

**Simulated vibrational coherence dynamics up to 2 ps**

To identify the vibrational mode underlying the observed photoelectron oscillations in Figure 3 and to determine whether its coherence survives after the $S_2 \rightarrow S_1$ transition, we performed a normal-mode projection analysis on the excited-state non-adiabatic dynamics trajectories from our previous study.[41] Snapshots from the excited-state trajectories were projected onto the normal modes of $S_0$ minimum. The details of this analysis are provided in the section S13 of the SI.

Normal-mode projection analysis identified distinct dynamical behaviors among the active vibrational modes during the nonadiabatic relaxation (see Figure S14 and S15 in section S13 of the SI). Our previous study showed that the ultrafast $S_2 \rightarrow S_1$ population transfer completes within approximately 30 fs and is mediated primarily by the BLA motion. Consistent with this mechanism, the two high-frequency stretching modes (Figure S14), namely the dominating C5=C6 stretch (1692 $cm^{-1}$) and the C4=O8 stretch (1710 $cm^{-1}$), constitute the BLA motion; they exhibit a single prominent oscillation within the early-time region (~30 fs), followed by an abrupt attenuation of their vibrational coherence upon reaching the $S_1$ state. In contrast, three strongly coherent lower-frequency modes (in-plane: 539, 599, and 736 $cm^{-1}$ shown in Figure 3c) display a much longer coherence lifetime (Figure S15). These modes, assigned to in-plane rocking motions (539 and 599 $cm^{-1}$) and the ring-breathing motion of pyrimidine ring (736 $cm^{-1}$), are activated in the $S_2$ state and retain their coherence through the conical intersection into the $S_1$ minimum. The comparison of the 736 $cm^{-1}$ ring-breathing mode with the observed photoelectron oscillations, and the origin of the measured phase difference, are taken up in the Discussion.

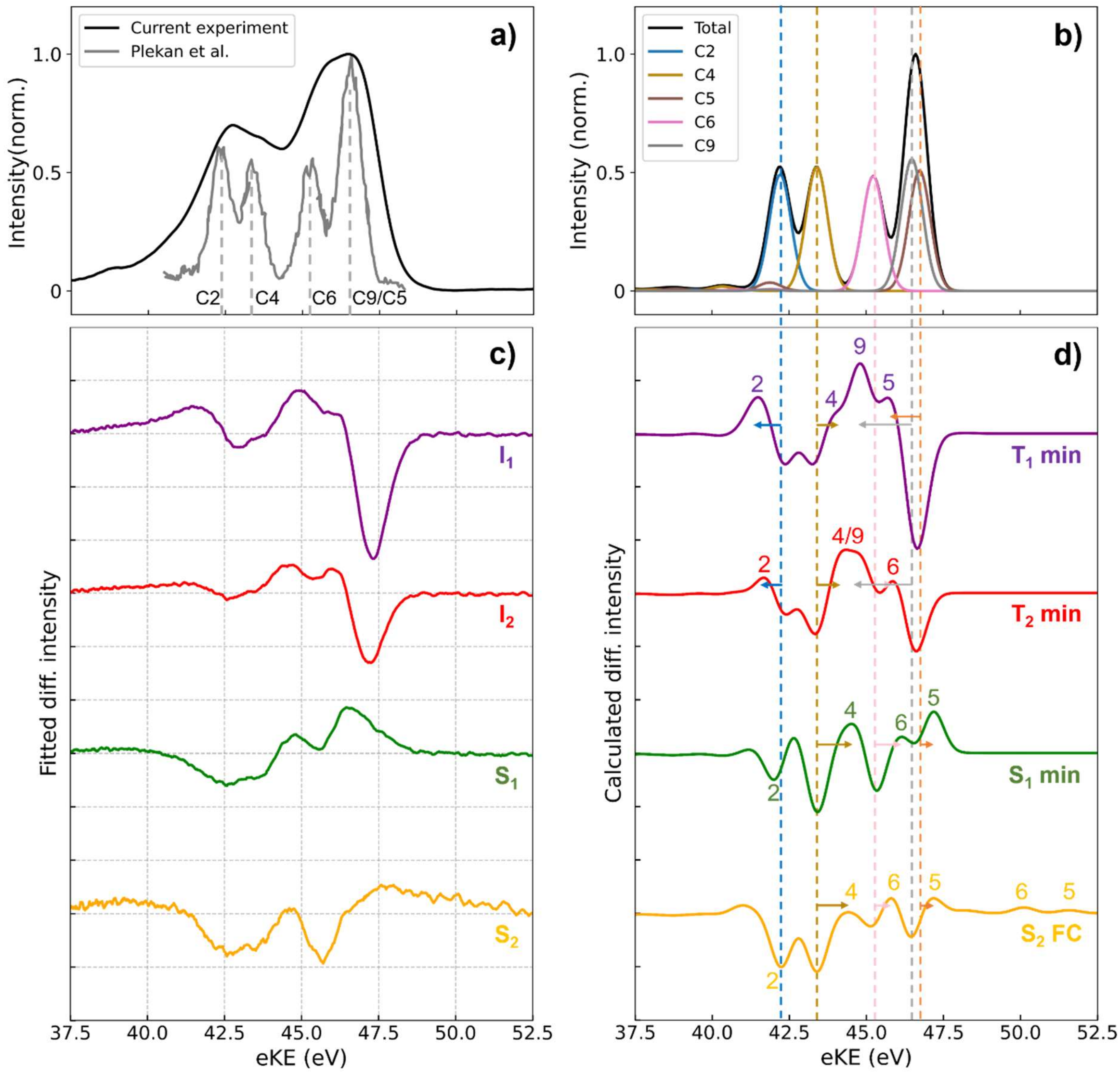

**Figure 5.** Comparison of the experimental and calculated carbon 1s XPS spectra of thymine. (a) Experimental static spectra of thymine. The ground-state spectrum from the present experiment is in black, and the atom-resolved static XPS spectrum from the synchrotron data is in grey. (b) Calculated static ($S_0$) XPS spectrum, with the contribution from each carbon atom in colored lines and the total in black. (c) State-associated difference spectra from the global fit, involving two singlet states ($S_2$, $S_1$) and two other intermediate states ($I_2$, $I_1$). (d) Calculated XPS difference spectra of each electronic state ($S_1$, $T_2$, $T_1$) at its minimum-energy geometry, together with $S_2$ state at FC geometry. Atomic contributions are indicated by carbon-atom number, and colored arrows mark the shift of each photoelectron peak relative to its ground-state XPS peak position.

## XPS difference spectrum simulation

To elucidate the relaxation pathway, we assigned the distinct features in the experimental difference spectrum by comparing them to calculated XPS spectra. Specifically, we computed the spectra for the relaxed electronic states ($S_0$, $S_1$, $T_2$, $T_1$) alongside the initially populated $S_2$ state at the FC geometry. Figure 5a presents the measured static spectrum from the present experiment alongside synchrotron data,[38] whose higher resolution enables the isolation of the photoelectron peaks of the individual carbon atoms in thymine. Figure 5b shows the calculated static spectrum decomposed into the contribution from each carbon atom. Consistent with the construction of the experimental difference spectra shown in Figure 5c (the state-associated spectra from the global analysis), the calculated difference spectra in Figure 5d were obtained by subtracting the ground-state ($S_0$) spectrum from that of each excited state.

For the initial $S_2$ state at the FC geometry, the calculated difference spectrum displays a bleach near 42.5 eV originating from a reduction in the C2 intensity together with a blue shift of the C4 contribution toward the higher eKE region. An intensity increase near 46.5 eV is primarily due to a blue shift of the C6 photoelectron peak, while the high-eKE tail (48.5–52.5 eV) arises mainly from the overlapping blue shifts of the C5 and C6 peaks. Meanwhile, the C9 contribution remains largely unchanged in both intensity and peak position, indicating a minor role in this spectral region. Moving towards the $S_1$ state, the theoretical difference spectrum closely resembles that of the $S_2$ state, having the same spin multiplicity (singlet). While the simulated $S_1$ XPS spectrum indicates a partial recovery of the C2 contribution at 42.5 eV compared to the $S_2$ XPS profile, it still yields a net negative signal in this region. Notably, the high-eKE tail characteristic of the $S_2$ at the FC region disappears at the $S_1$ minimum.

Turning to the triplet manifold (Figure 5d), the simulated $T_2$ difference spectrum exhibits a UV-induced intensity increase at 41–42 eV arising from a red shift of the C2 photoelectron peak—a feature absent in $S_1$. The blue shift of the C4 peak is much smaller in $T_2$ than in $S_1$, which reduces the negative feature at 43.5 eV. Most importantly, a ~2 eV red shift of the C9 peak produces a dramatic negative feature near 47 eV; this shift, combined with a blue shift of the C6 peak, gives rise to the UV-induced increase at 44–46 eV. The calculated $T_1$ difference spectrum closely mirrors that of $T_2$, with two notable differences. First, the negative feature at 47 eV becomes more intense in $T_1$ because both the C9 and C5 peaks red-shift, in contrast to $T_2$ where only the C9 peak shifts. Second, the UV-induced increase at 41–42 eV is stronger in $T_1$, reflecting a larger red shift of the C2 peak than in $T_2$.

## Discussion

### Comparison of XPS simulations with the experiments and special role of C9 in identifying the molecular dynamics

Based on the high-eKE tail above 48.5 eV, the fact that it is the first observable feature and its short lifetime, we assign R1 to the $S_2$ ($^1\pi\pi^*$) electronic state. In addition, based on the 6.3 ps trapping lifetime in R2, we assign region R2 to the $S_1$ ($^1n\pi^*$) electronic state. This assignment aligns with previous gas-phase studies[10,14,15] and is strongly supported by our simulated XPS difference spectra. For R1, the calculated $S_2$ FC spectrum (Figure 5d, yellow line) successfully reproduces both the 42.5 eV bleach and the high-eKE tail, driven by spectral shifts of the ring carbon atoms (C2, C4, C5, C6). For R2, the $S_1$ minimum simulation captures the analogous 42.5 eV bleach and the intensity increase between 46–47.5 eV, arising from these same four carbon atoms. Furthermore, the simulation correctly predicts the disappearance of the high-eKE tail upon relaxation to the $S_1$ state. The experimental features around 45–46 eV in R1 and R2 are not fully reproduced by the simulations. The disappearance of the bleach feature at 45.5 eV as thymine relaxes from $S_2$ to $S_1$ may originate from the shift of the C6 photoelectron line, which depends strongly on the C5=C6 bond length modulation and therefore differs between the $^1\pi\pi^*$ and $^1n\pi^*$ states.

Notably, our simulations predict no measurable shift or intensity change for the C9 peak across both singlet states (see Figure 5d). This predicted spectral inactivity directly confirms that the methyl group ($C9H_3$) does not participate electronically in the $S_2 \rightarrow S_1$ internal conversion, a conclusion made possible by the site selectivity of XPS. These results reveal that the methyl group plays a predominantly inertial role during the initial singlet-state relaxation, rather than directly modifying the electronic character of the excited states. Although direct $^1\pi\pi^* \rightarrow S_0$ relaxation appears feasible due to a nearly barrierless pathway on the potential energy surface (see Figure S17 in Section S16 of the SI), the increased inertia introduced by the methyl group hinders the ultrafast C5 pyramidalization required for this transition.

Comparing the simulations with the experimental data for R3 and R4, from which the experimental $I_2$ and $I_1$ spectra are respectively extracted, the simulated XPS difference spectra for the $T_2$ and $T_1$ states show excellent overall agreement with the experimental $I_2$ and $I_1$ spectra extracted from regions R3 and R4, respectively. Specifically, the UV-induced signal increase below 42 eV and the adjacent bleach around 42.5 eV originate predominantly from C2 shifts. The intensity rise near 45 eV arises from shifts in C4, C6, and C9 for the $T_2$ ($^3n\pi^*$) state, and from C4, C5, and C9 for the $T_1$ ($^3\pi\pi^*$) state. These spectral signatures allow us to securely assign the $I_1$ spectrum, derived from R4, to the $T_1$ state; notably, the C9 atom plays a critical role in driving the dramatic hallmark bleach near 47.5 eV. While the $I_2$ spectrum, derived from R3, strongly resembles the simulated $T_2$ state, we postpone its formal assignment to incorporate further arguments based on potential energy curve simulations.

This prominent involvement of the C9 atom in the triplet manifold represents a stark reversal from the singlet-state dynamics. The methyl group, which acts primarily as an inertial spectator during singlet-state relaxation, emerges upon triplet formation as a highly sensitive reporter of intersystem crossing. In the calculated spectra at the $T_2$ and $T_1$ minima, the C9 photoelectron peak exhibits a pronounced red shift to ~45 eV in the eKE region. Thus, the positive signal at ~47.5 eV characteristic of the $S_1$ minimum is inverted, producing the dramatic hallmark bleach alongside a substantial intensity increase near 45 eV—a signature entirely consistent with the experimental spectrum.

The singlet-triplet sensitivity of the C9 shift reflects a reorganization of hyperconjugative (σ-π) interactions between the methyl group and the heterocycle upon entering the triplet manifold. This state-dependent modulation of hyperconjugation across the different electronic states is clearly visible in the potential energy curve along the $-CH_3$ rotational coordinate (see Figure S18 in section S17 of the SI). The ground state exhibits a well-defined torsional minimum at -59°, indicating strong hyperconjugation. This barrier is absent in the $S_1$ ($^1n\pi^*$) state, pointing to negligible hyperconjugation and nearly free methyl rotation. In the $T_1$ ($^3\pi\pi^*$) state, however, hyperconjugation is not only restored but effectively reversed with respect to $S_0$, shifting the torsional energy minimum to -5°. Attachment/detachment density analysis confirms this state-dependent electronic structure, revealing a significantly larger electron density transfer from the local σ orbitals of the $-CH_3$ group to the ring's π-system in the $T_1$ state than in the $S_1$ state (see Figure

S18 in section S17 of the SI). This enhanced, geometry-dependent charge transfer through C5 drives a prominent blue shift of the C9 1s binding energy, governed by the strength of the σ-π orbital mixing.[43] The C9 photoelectron signal therefore serves as a highly sensitive, site-specific signature of intersystem crossing and triplet state formation, aligning perfectly with the concept of an excited-state chemical shift.[28]

While the preceding analysis focuses on initial-state charge distributions, final-state effects following photoionization also fundamentally shape the XPS signatures. Specifically, the differing spatial symmetries of the singlet and triplet wavefunctions directly impact core-hole screening. Because the triplet state possesses a symmetric spin wavefunction with respect to particle exchange, its spatial wavefunction is rigorously anti-symmetric. The resulting exchange interaction forces valence electrons to spatially avoid one another in addition to standard Coulomb repulsion, rendering them less effective at screening the localized core hole than in the corresponding singlet state. This less efficient screening shifts the triplet photoelectron peaks toward lower kinetic energies. Supporting this physical picture, simply shifting the ground-state spectrum to lower kinetic energies produces a difference spectrum that closely mimics the experimental triplet signal (see Figure S5 in section S4 of the SI). Although our theoretical XPS simulations inherently encompass these final-state effects, this symmetry-driven difference in screening provides a powerful, intuitive rationale for the global kinetic energy shifts observed upon triplet formation.

### Structural dynamics resolved by vibrational coherence via tr-mc-XPS

The multicenter character of the XPS signal provides a second dimension of mechanistic information by linking changes in local electronic structure to coherent nuclear motion. Our XPS difference spectrum exhibits pronounced photoelectron-intensity oscillations at ~730 $cm^{-1}$, with a phase difference of π between the bleach and the UV-induced signal increase regions (Figure 3). This demonstrates vibrational coherence generated by $^1\pi\pi^*$ excitation and transferred to the $^1n\pi^*$ state within ~200 fs. The normal-mode projection analysis supports this interpretation: the simulated 736 $cm^{-1}$ ring-breathing mode persists through the $S_2$ ($^1\pi\pi^*$) → $S_1$ ($^1n\pi^*$) internal conversion and remains coherent up to ~750 fs, matching the observed oscillation. The two other low-frequency modes predicted at 539 and 599 $cm^{-1}$ are not clearly resolved experimentally, most likely because of insufficient temporal sampling at longer delays combined with their weak intensity. The high-frequency modes at 1692 and 1710 $cm^{-1}$, which drive the internal conversion, likewise remain unresolved, and we are not able to judge if the coherence is transferred in the CI due to the limited temporal resolution of the measurement.

Karashima et al. also observed this ring-breathing mode at 750 $cm^{-1}$ in the thymine and uracil relaxation dynamics in the gas phase using tr-EUV-PES.[23] The observed vibrational coherence is transferred to the $^1n\pi^*$ state, consistent with our observation for thymine using tr-mc-XPS. Faccialà et al., however, did not identify the 750 $cm^{-1}$ frequency in gas-phase uracil using time-resolved XPS at the O 1s core level.[31] This difference likely reflects the distinct site sensitivity of the measurements: our carbon XPS approach probes multiple endocyclic atomic centers, whereas their experiment primarily monitors exocyclic oxygen, rendering it less sensitive to the skeletal ring-breathing modes.

The advantage of the multi-center probing manifests itself in the additional relative phase information of the oscillations at different atomic sites. We analyzed site-specific atomic contributions using theoretical calculations to elucidate the π phase shift between the bleach and UV-induced signal increase regions shown in Figure 3b. Based on the site-resolved difference spectra in Figure 5d, which indicate that C2, C4, and C6 dominate these difference-spectral features in $S_1$ minimum, we simulated the XPS spectra by considering only contributions from these atoms. By mapping these contributions along the ring-breathing vibrational coordinate (see Figure S13 in section S12 of the SI), we found that nuclear displacements modulate the ionization probability of C2 with an opposite phase relative to that of C4 and C6, directly reproducing the π phase shift observed between the bleach and UV-induced spectral regions. Upon ring expansion, the XPS intensity of C2 is enhanced, whereas those of C4 and C6 are attenuated. In contrast, ring contraction reduces the C2 intensity while slightly enhancing or preserving the XPS intensity of C4 and C6.

This site-specific intensity modulation originates from the distinct response of the core-ionized cationic states to nuclear deformation (final-state effect), which can be rationalized from the spatial landscape of the molecular orbitals (Figure S13). The C4 and C6 sites belong to the same π-conjugated system, extending across the main C6=C5–C4=O8 moiety. Consequently, photoionization at either site produces a cationic state in which the resulting final state is delocalized over this conjugated network, leaving the valence π and π* orbitals largely unchanged for the cation compared to the neutral state. In contrast, the C2 carbonyl group forms a local π-system that is orthogonal to the C4, C5 and C6 π-network. Photoionization at the C2 therefore produces cationic states with electron density localized at the C2 carbonyl group. The corresponding relaxed valence orbitals display a nodal topology that is approximately rotated by 90° shift relative to that of the C4/C6 cationic states (see details in section S12 of SI). As a result, changes in the bond length alternation along the ring-breathing coordinate affect the two types of cationic states in opposite ways: ring expansion stabilizes one configuration while destabilizing the other, and vice versa for ring contraction. This opposite energetic response modulates the C2 and C4/C6 ionization intensities with opposite phases, ultimately giving rise to the observed 180° (π) phase shift in the oscillatory XPS signal.

## Assignment of $I_2$ and $I_1$

The experimental spectra associated with intermediates $I_2$ and $I_1$ (Figure 5c), which correspond respectively to regions R3 and R4 of the difference spectrum (Figure 2b), exhibit a pronounced bleach at an eKE of ~47 eV, distinguishing them from $S_2$ ($^1\pi\pi^*$) and $S_1$ ($^1n\pi^*$). The difference spectra of $I_2$ and $I_1$ resemble the calculated difference spectra of $T_2$ ($^3n\pi^*$) and $T_1$ ($^3\pi\pi^*$), respectively (Figure 5d), pointing to either the population of triplet states, or a mixed signal stemming from both triplet and singlet electronic states. Our earlier theoretical analysis of the $S_1/T_2/T_1$ crossing region establishes that favorable spin-orbit coupling heavily biases the system toward direct $S_1 \rightarrow T_1$ transfer. Consequently, $T_2$ either does not participate in a sequential relaxation cascade $S_1 \rightarrow T_2 \rightarrow T_1$ or is populated only transiently before rapid internal conversion to $T_1$. In either case, $T_2$ cannot be assigned to the long-lived $I_2$ or $I_1$ signals, leaving $T_1$ as the sole triplet state capable of trapping the electronic population on the experimental timescale.

If we set aside the possibility of a mixed singlet–triplet signal in the assignment of $I_2$, the only remaining option is to attribute both $I_2$ and $I_1$ to $T_1$, namely the hot $T_1$ for $I_2$ and the $T_1$ min for $I_1$. This assignment, however, is inconsistent with our calculations, showing that the XPS spectrum of hot $T_1$ is essentially a broadened version of the $T_1$-minimum spectrum, with a 47.5 eV bleach of comparable amplitude in the two cases (see the hot-$T_1$ XPS difference spectrum in Figure S11 in section S10 of the SI). This cannot account for the measured 47.5 eV bleach of $I_2$, which is only about half as strong as that of $I_1$. We therefore exclude the assignment of $I_2$ and $I_1$ to hot $T_1$ and the $T_1$ minimum, respectively.

We further exclude the hot ground state as a possible explanation for $I_2$ as a 1.6 eV energy gap between $CI_{10}$ ($^1n\pi^*/S_0$) and $S_1$ minimum hinders the nonradiative $S_1 \rightarrow S_0$ internal conversion pathway. It is therefore reasonable to propose that the spectrum of $I_2$ arises from a superposition of the singlet state, here called $S_1^*$, and the triplet state $T_1$, whereas the spectrum of $I_1$ corresponds to that of $T_1$ alone, consistent with the expected singlet-to-triplet relaxation sequence. Here $S_1^*$ denotes a further-relaxed form of $S_1$ and has a feature ~47 eV that is less negative or even positive to yield a less negative feature when it is combined with $T_1$, which could match the less negative feature of $I_2$ compared to that of $I_1$.

Crucially, the $S_1^*$ spectrum cannot be identified with the $S_1$ spectrum shown in Figure 5c, as the failure of the three-state model in the explanation of the data shows (see section S3 of the SI). Because $S_1$ and $T_1$ are treated as independent components there, the fit already spans any linear combination of their spectra; the persistent, systematic residuals therefore point to a distinct component that the model does not capture. We also verified that these residuals are physically real, rather than artifacts resulting from instrumental drift (such as shifts in the X-ray photon energy during data acquisition). Instead, they are induced by the UV excitation itself, reflecting the change in the molecule's own electronic population.

To test the assignment of $I_2$ to mixture of $S_1^*$ and $T_1$, we extended the cascade model with a branch $S_1 \rightarrow S_1^* \rightarrow I_1$ alongside the main relaxation cascade $S_2 \rightarrow S_1 \rightarrow I_1$. In this extended model, $I_2$ no longer appears as an independent species: the population previously described as $I_2$ corresponds to the coexisting $S_1^*$ and $I_1$ populations, whose combined spectrum the four-state model returned as a single component. In the higher eKE region, the $S_1^*$ spectrum from the global analysis (see details in section S5 of SI) closely resembles the calculated hot-$S_1$ (vibrationally excited) spectrum (see Figure S6 in section S5 of SI). However, this correspondence degrades at lower eKE, preventing a definitive assignment to the hot $S_1$ state without further experimental and theoretical investigation beyond the scope of this study. Its extracted lifetime, (240 ± 28) ps, agrees well with the $I_2$ lifetime of (239 ± 29) ps. Together these support the assignment of the $I_2$ spectrum to a mixture of $S_1^*$ and $T_1$. Independent support comes from gas-phase tr-PES on thymine, where Miura et al. report a triplet yield of ~86% from $S_1$.[10] The apparent yield deficit is therefore precisely what our model predicts, and it explains why the channel has not previously been resolved.

We assigned the $I_1$ spectrum to $T_1$ ($^3\pi\pi^*$) solely based on the fact that the simulated $T_1$ XPS difference spectrum reproduces the features observed in the I1 spectrum. To verify that $T_1$ is the final state populated within our measurement window, we tested an alternative explanation for this long-lived state: hot ground state, which is possibly the only relaxation product of $T_1$. The simulated hot ground state difference spectrum, however, does not reproduce the fitted $I_1$ spectrum (see Figure S11 in section S10 of the SI for more details). Furthermore, the calculated potential energy curve rules out the twisted $S_0$ intermediate, as the low barrier back to the planar $S_0$ minimum (Figure S12 in section S11 of the SI). These results leave $T_1$ as the only plausible candidate for $I_1$, associated with R4. This assignment is consistent with earlier work by Ligare et al. and Yu et al.,[19,21] who attributed long-lived states with lifetimes beyond 250 ns and beyond 1 ns, respectively, to $T_1$ ($^3\pi\pi^*$).

## Conclusion

In the present study, we investigated the electronic and structural relaxation from $^1\pi\pi^*$ state of thymine in the gas phase using tr-mc-XPS, supported by state-of-the-art *ab initio* quantum calculations. The multi-center nature of tr-mc-XPS yields a rich, well-structured time-resolved difference spectrum, enabling a comprehensive study of the ultrafast $^1\pi\pi^*/^1n\pi^*$ IC pathway and the subsequent ISC to the long-lived 'dark state'. Crucially, we reveal that this 'dark state' is not exclusively a triplet state. For the first 240 ps following ISC, a mixture of the $S_1^*$ singlet state and the $^3\pi\pi^*$ triplet state dominates the dynamics before transitioning to a purely $^3\pi\pi^*$ triplet state for the remainder of the 1 ns measurement window. This assignment is rationalized by the calculation of different spectra in different excited states and potential energy curve calculations. Specially, the optimization of the $^1n\pi^*/^3n\pi^*/^3\pi\pi^*$ crossings clearly demonstrates that the $^1n\pi^*/^3\pi\pi^*$ ISC does not require the $^3n\pi^*$ as an intermediate bridge.

Alongside these electronic population dynamics, we also captured the ultrafast coherent nuclear motion that accompanies the initial excitation. We observed pronounced photoelectron-intensity oscillations at ~730 $cm^{-1}$, exhibiting a phase difference of π between the bleach and the UV-induced-increase regions. This oscillation corresponds to the ring-breathing normal mode which is initiated by the $^1\pi\pi^*$ excitation and survives the IC to $^1n\pi^*$ state within ~200 fs, maintaining coherence up to ~750 fs. The π phase shift originates from the site-specific, opposing responses of the core-ionized cationic states to nuclear deformation during ring contraction and expansion. Observing this vibrational coherence highlights the advantage of the multi-center probing capability of tr-mc-XPS, which can map complex nuclear dynamics across distinct atomic sites of the thymine pyrimidine ring.

Finally, we elucidate the dual role of the methyl group during the IC and ISC of thymine. During the initial singlet relaxation, the methyl group acts strictly as an inertia spectator, exhibiting no electronic participation in the $^1\pi\pi^* \rightarrow {}^1n\pi^*$ IC. From a structural point of view, instead, its mass hinders the ultrafast C5 pyramidalization required for direct $^1\pi\pi^* \rightarrow S_0$ relaxation, as observed in uracil. Remarkably, this role is reversed upon triplet formation, where the methyl group becomes a highly sensitive electronic reporter. State-dependent hyperconjugation between the local σ orbitals of the methyl group and the ring's π-system in the $^3\pi\pi^*$ state drives the electron density from C9 to the pyrimidine ring. This charge shift results in a prominent blue-shift of the C9 1s electron binding energy, directly causing the pronounced bleach feature observed at ~ 47.5 eV eKE region.

While our current tr-mc-XPS measurements and simulations successfully resolve these primary relaxation pathways, the precise structural and electronic evolution of the intermediate $S_1^*$ state during the 240 ps mixed-state window warrants further investigation. A previous time-resolved X-ray absorption spectroscopy (tr-XAS) study has successfully probed thymine dynamics up to 30 ps.[14] However, to fully isolate the $S_1^*$ characteristics from the $^3\pi\pi^*$ background, extending Oxygen K-edge tr-XAS or tr-XPS to significantly longer time delays—matching the extended temporal range accessed in our work—will be crucial. Because the $S_1^*$ state possesses $n\pi^*$ character, this approach will provide enhanced, site-specific sensitivity to the transient dynamics at the carbonyl centers by directly probing the involved oxygen lone-pair orbitals. Future theoretical work will need to connect the early-time non-adiabatic relaxation to the slower structural and electronic evolution observed at longer delays, to establish how the initial dynamics shape the population trapped in $S_1^*$. Ultimately, these complementary approaches will build upon the multi-center capabilities demonstrated here to complete our understanding of the multi-timescale photodynamics governing nucleobase stability.

## Acknowledgement

We acknowledge funding by the Deutsche Forschungsgemeinschaft (DFG) via Grant GU-1478/1-1. We acknowledge DESY (Hamburg, Germany), a member of the Helmholtz Association HGF, for the provision of experimental facilities. Part of this research was carried out at FLASH2 at DESY. Beamtime was allocated for proposal F-20231449. Z.P. was supported by the Chemical Sciences, Geosciences, and Biosciences Division, Office of Basic Energy Sciences, Office of Science, U.S. Department of Energy, through grant no. DE-FG02-86ER13491. M.H.R. and C.H.C. acknowledge support by the "PHC Star" program funded by the French Ministry for Europe and Foreign Affairs, the French Ministry for Higher Education, Research and Space, and the National Research Foundation of Korea - NRF. Centre de Calcul Intensif d'Aix-Marseille is acknowledged for granting access to its high-performance computing resources. The authors used AI to improve the readability and language of the manuscript. The authors reviewed and edited the output and took full responsibility for the content.

## Author contribution

M.G. and X.W. conceived the project and wrote the beamtime proposal. X.W. coordinated the experiment and the discussions between theory and experiment, and prepared the sample. G.C., F.P., and S.S. carried out the laser preparation. X.W., D.M., F.L., G.C., M.D.F., B.E., M.K., Z.P., O.P., F.P., S.S., and M.G. performed the experiment. X.W. analyzed, curated, and visualized the experimental data, and X.W., M.G., D.M., and F.L. validated the data. X.W. and W.P. interpreted the experimental data on the basis of simulations. W.P. produced, analyzed, curated, and visualized the theoretical data, and W.P. and M.H.-R. validated the data. X.W. wrote the original draft with input from W.P., M.G., C.H.C., and M.H.-R. X.W., M.G., W.P., and M.H.-R. discussed and revised the manuscript into its final form. All authors reviewed and edited the manuscript. M.G. supervised the experimental group. M.G., M.H.-R., and C.H.C. acquired funding. All authors have given approval to the final version of the manuscript.

# Supporting information

**S1. The experimental scheme for time-resolved X-ray photoelectron spectroscopy (XPS)**

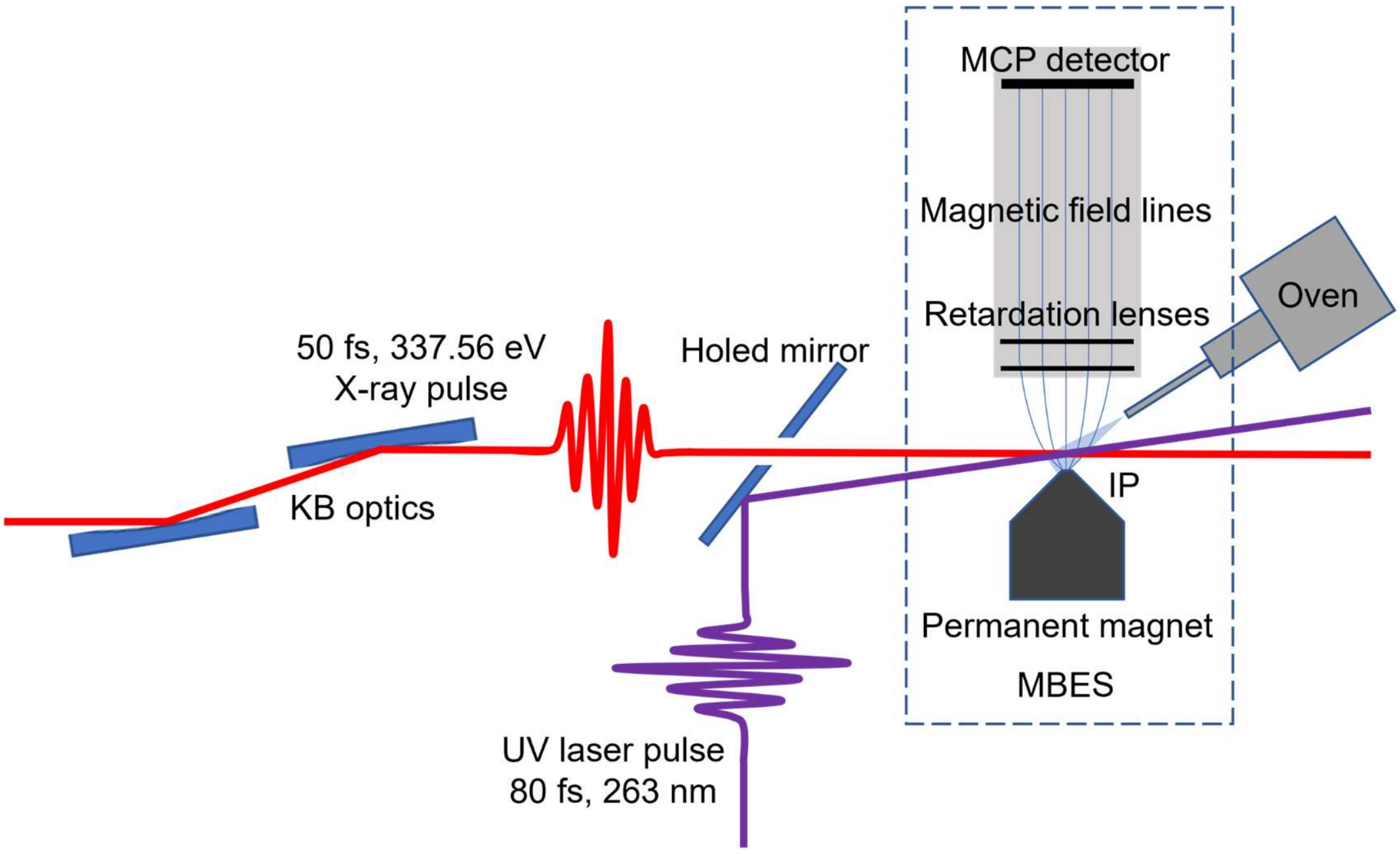

**Figure S1**. We performed time-resolved UV-pump X-ray-probe photoelectron spectroscopy on isolated thymine molecules at the carbon 1s core levels. The thymine molecules are evaporated from an oven (heated to 180 °C) and guided to the interaction point (IP) through an aluminum capillary. The 50 fs X-ray pulse with a photon energy of 337.56 eV is focused by the KB optics onto the interaction point inside the sample chamber, where the photoelectrons ejected from the sample are collected by a magnetic bottle electron spectrometer (MBES). A mirror with a hole (labeled 'holed mirror' in the plot) guides the incident 80 fs, 263 nm UV beam to the interaction region while letting the X-ray pulse pass through.

## S2. Time zero finding and estimation of the instrument response function (IRF)

In the data analysis, we first corrected the X-ray pulse arrival times for temporal jitter (on the order of a few hundred femtoseconds) relative to the nominal delays set during data acquisition. The jitter-corrected delays were then rebinned to give approximately equal statistics per bin. Both steps were applied only to delays within 1.1 ps, where the jitter is larger than the delay step and therefore limits the temporal resolution; at longer delays the step size is large compared with the jitter, so neither correction nor rebinning was necessary. Figure S2a shows the thymine carbon 1s difference spectrum obtained after this procedure. To extract time zero — the point of maximal temporal overlap between the UV pump and X-ray probe pulses — the intensity integrated along the electron kinetic energy (eKE) axis was fitted as a function of pump–probe delay using a modified error function. This integration was carried out in two steps: first, the absolute value of the 2D difference spectrum was taken, since the spectrum contains both positive and negative features whose contributions would otherwise cancel upon direct integration; the rectified spectrum was then integrated along the eKE axis to yield the delay-dependent intensity. The modified error function results from the convolution of the instrument response function (IRF) — taken as a Gaussian of standard deviation of $\sigma$ — with a Heaviside step function, and has the form

$$S(t) = \frac{1}{2}\left[1 + erf\left(\frac{t-\mu}{\sigma\sqrt{2}}\right)\right]$$

where the error function is defined as

$$erf(z) = \frac{2}{\sqrt{\pi}}\int_0^z e^{-t^2}\, dt$$

Here $t$ is the pump-probe delay, $\mu$ is the delay at which the X-ray probe and UV pump pulses overlap, and $\sigma$ is the width of the IRF. Figure S2b shows the integrated intensity in the eKE range 41–48.5 eV as a function of delay (blue) together with the fit (red), which yields $\mu$ = (-0.002 ± 0.005) ps and $\sigma$ = (0.067 ± 0.005) ps, corresponding to an IRF of (158 ± 12) fs at full width at half maximum (FWHM). Uncertainties in both parameters were estimated by bootstrap resampling.

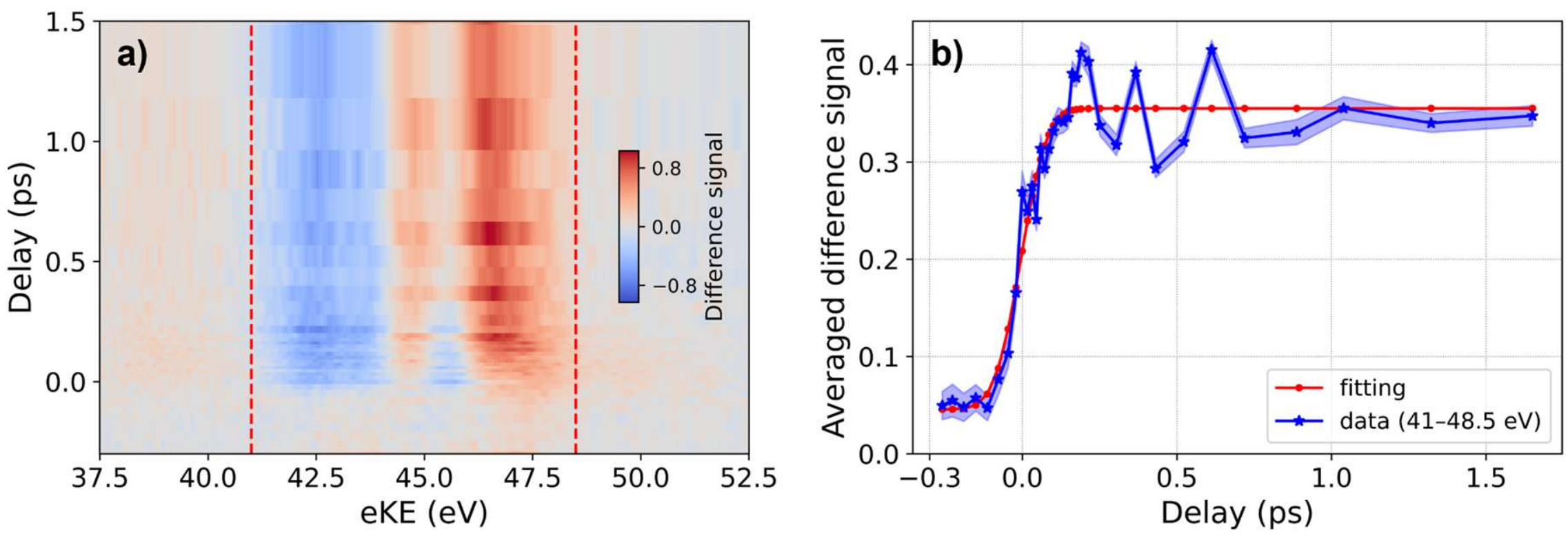

**Figure S2.** (a) Time-resolved carbon 1s XPS difference spectrum of gas-phase thymine measured in the present experiment. Delays within the first 1.1 ps have been corrected for X-ray arrival-time jitter to recover the highest possible temporal resolution; longer delays are shown as recorded. Red dashed lines mark the eKE region used for integration. (b) Intensity integrated along the eKE axis over 41–48.5 eV (blue) together with a fit to a modified error function (red). Prior to integration, the absolute value of the 2D difference spectrum was taken to prevent cancellation between its positive and negative features. Error bars were estimated by bootstrap resampling. The fit yields $\mu = (-0.002 \pm 0.005)$ ps and $\sigma = (0.067 \pm 0.005)$ ps; see the main text for details.

## S3. Global analysis: three-state & four-state kinetic model

We used a global analysis method to extract the dynamic populations and state-associated spectra from the experimental XPS difference spectrum, which is a two-dimensional spectrum $N_{delay} \times N_{eKE}$ ($N_{delay}$ is the number of the pump–probe delay points and $N_{eKE}$ is the number of eKE bins). In this approach, a single kinetic model describes the transient behavior at all eKE simultaneously. The global fitting method enables us to estimate the lifetimes of the involved electronic states and facilitates the direct comparison between the calculated difference spectrum of each electronic state and the corresponding state-associated spectrum. The dynamic populations are determined by the kinetic model as the solutions of a set of coupled ordinary differential equations, whose rate parameters are the inverse lifetimes of the involved physical states. From the parameterized dynamic populations, the corresponding state-associated difference spectra are obtained from the experimental difference spectrum using the Moore–Penrose pseudoinverse. Multiplying the dynamic populations by the state-associated spectra yields the modeled difference spectrum. By minimizing the weighted least square $\chi^2$, we obtain the optimized rate parameters as well as the optimized dynamic population profiles and state-associated spectra.

The key step in the global analysis is the construction of the kinetic model describing the relaxation of gas-phase photoexcited thymine. From previous studies,[1–3] at least three states are known to be involved in the relaxation dynamics: the initially populated excited state $S_2$ ($^1\pi\pi^*$); $S_1$ ($^1n\pi^*$), reached via a conical intersection; and finally, a long-lived dark state $I_1$, whose character this work aims to clarify in the main text. We therefore first test the minimal cascade kinetic model: $S_2 \rightarrow S_1 \rightarrow I_1$. The instrument response function is modeled as a Gaussian with a width of $\sigma$, which has the form

$$g(t) = \frac{1}{\sqrt{2\pi}\sigma} exp\left(-\frac{t^2}{2\sigma^2}\right)$$

where $\sigma$ is treated as a fit parameter and optimized together with the rate constants $k_1$, $k_2$ introduced below. Rather than convolving the populations with the IRF after solving the rate equations, we include the IRF directly as a source term in the coupled ordinary differential equations — an equivalent treatment for linear kinetics:

$$\frac{dS_2}{dt} = g(t) - k_1 S_2$$

$$\frac{dS_1}{dt} = k_1 S_2 - k_2 S_1$$

$$\frac{dI_1}{dt} = k_2 S_1$$

where $k_1$, $k_2$ are rate constant for the population of $S_2$ and $S_1$, respectively. The coupled ordinary differential equations are solved numerically using the odeint routine from the scipy.integrate module to obtain the population dynamics. The rate constants $k_1$, $k_2$, together with the IRF width $\sigma$, are optimized by minimizing $\chi^2$ using scipy.optimize.minimize with the 'Nelder–Mead' method.

$\chi^2$ is defined as

$$\chi^2 = \sum_{i=1}^{N_{eKE}} \sum_{j=1}^{N_{delay}} \frac{[S_{mod}(\varepsilon_i, t_j) - S_{exp}(\varepsilon_i, t_j)]^2}{u^2(\varepsilon_i, t_j)}$$

where:

- $\varepsilon_i$ is the $i$-th electron kinetic energy bin, $i = 1, \dots, N_{eKE}$
- $t_j$ is the $j$-th pump–probe delay, $j = 1, \dots, N_{delay}$
- $S_{mod}$ and $S_{exp}$ are the modelled and experimental difference spectra
- $u(\varepsilon_i, t_j)$ is the uncertainty of the experimental value at that point, which is estimated by the bootstrap resampling on the experimental data.

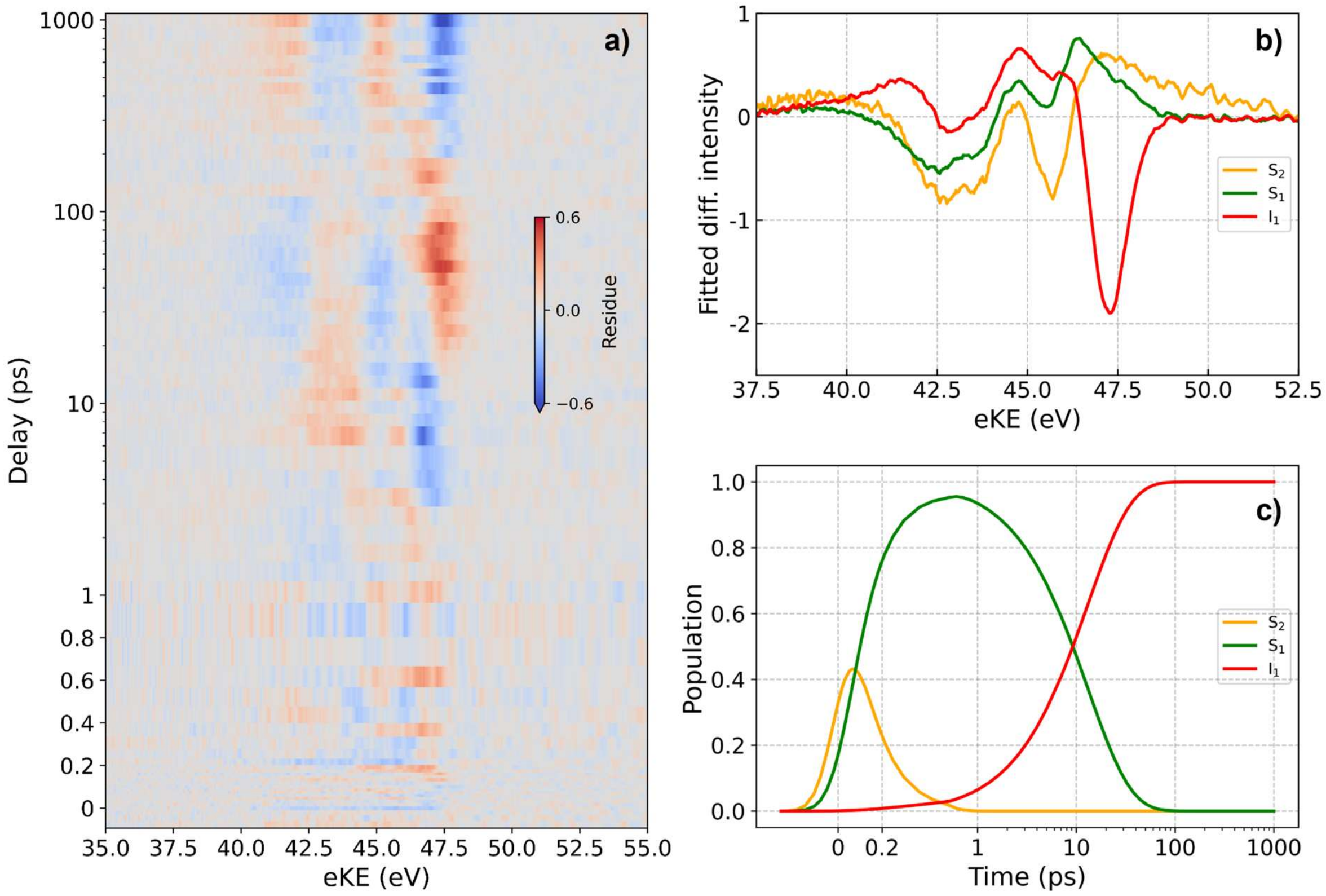

**Figure S3.** (a) Residual, defined as the difference between experimental difference spectrum and the optimized fit, from the three-state model, with a systematic deviation clearly visible beyond 5 ps. (b) State-associated spectra corresponding to the involved states: $S_2$, $S_1$, $I_1$. (c) Dynamic populations of the involved states. The optimized fit yields lifetimes of (122 ± 33) fs and (13.1 ± 0.8) ps for the $S_2$ and $S_1$ states, respectively, and an IRF with a standard deviation of (73 ± 5) fs.

The optimized parameters yield lifetimes of (122 ± 33) fs and (13.1 ± 0.8) ps for the $S_2$ and $S_1$ states, respectively, and an IRF with a standard deviation of (73 ± 5) fs, comparable to the value estimated in section S2 of the SI. Although the optimization terminates successfully, the reduced chi-squared — defined as $\chi_r^2 = \chi^2/(N-3)$ with $N = N_{delay} \times N_{eKE}$ — takes a value of 1.73, noticeably larger than 1, indicating imperfect agreement between the model and experimental data. Figure S3 shows the residual, state-associated spectra, and dynamic populations from the optimized fit. A systematic deviation is clearly visible beyond 5 ps in the residual (Figure S3a), where the largest amplitude reaches roughly one-third of that of the experimental difference spectrum. We confirmed that this residual is real — not caused by a drift in the X-ray photon energy during data acquisition, but induced by the UV excitation itself, i.e. by the change in the molecule's own electronic population. Moreover, the fitted $S_1$ lifetime is substantially longer than the ~6 ps value widely reported in previous studies.[3–6] Together, these observations indicate that the three-state kinetic model cannot accurately describe the relaxation of photoexcited thymine in the gas phase. We therefore turn to a four-state model, $S_2 \rightarrow S_1 \rightarrow I_2 \rightarrow I_1$, which introduces an additional intermediate state $I_2$ with rate constant $k_3$, described by the following coupled differential equations:

$$\frac{dS_2}{dt} = g(t) - k_1 S_2$$

$$\frac{dS_1}{dt} = k_1 S_2 - k_2 S_1$$

$$\frac{dI_2}{dt} = k_2 S_1 - k_3 I_2$$

$$\frac{dI_1}{dt} = k_3 I_2$$

The optimization again terminates successfully, now with $\chi_r^2 = 0.95$, very close to one, indicating excellent agreement between the model and the experimental data. Introducing the additional intermediate state reduces the $\chi^2$ from 90693 to 50042, a 45% improvement obtained at the cost of a single additional parameter. According to the optimized four-state fit, the extracted lifetimes are (92 ± 18) fs for $S_2$ and (6.3 ± 0.2) ps for $S_1$. Notably, the $S_1$ lifetime obtained from the four-state model agrees well with the ~6 ps value reported previously, in contrast to the anomalously long value by the three-state fit. $I_2$ exhibits a lifetime of (239 ± 29) ps, while $I_1$ persists beyond 1 ns, exceeding the experimental time window. The instrument response function is characterized by a Gaussian with a standard deviation of (72 ± 5) fs. Figure S4 shows the residual, state-associated spectra, and dynamic populations from the four-state kinetic model. Crucially, the systematic deviation that dominated the three-state residual is entirely absent: the bipolar spectral pattern spanning ~41–48.5 eV, together with its sign reversal between ~30–100 ps and >300 ps, is removed, leaving only incoherent scatter about zero across the full delay range (Figure S4a, plotted on the same color scale as Figure S3a). The timescale of this previously unaccounted-for component is consistent with the fitted $I_2$ lifetime of (239 ± 29) ps, confirming that the three-state model was missing a genuine kinetic channel rather than suffering from an instrumental artifact. Uncertainties were estimated by bootstrap resampling combined with the global fit.

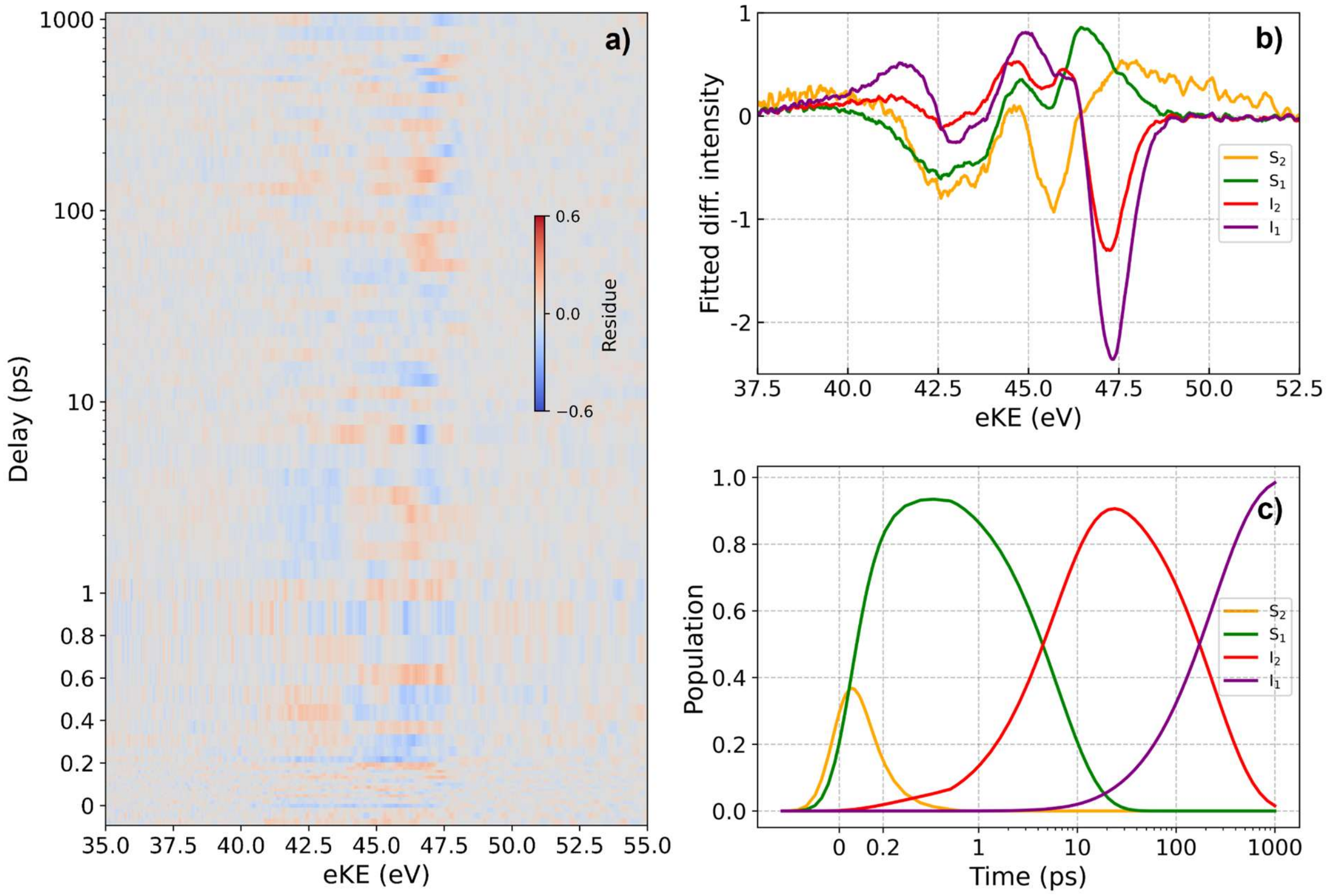

**Figure S4.** (a) Residual from the four-state kinetic model, with a systematic deviation clearly removed compared to that of the three-state model. (b) State-associated spectra corresponding to the involved four states. (c) Dynamic populations of the involved states. The optimized fit yields lifetimes of (92 ± 18) fs for the $S_2$ state, (6.3 ± 0.2) ps for the $S_1$ state, (239 ± 29) ps for the additional intermediate state $I_2$, and an IRF with a standard deviation of (72 ± 5) fs.

## S4. Distinguishing singlet and triplet states via symmetry-driven core-hole screening

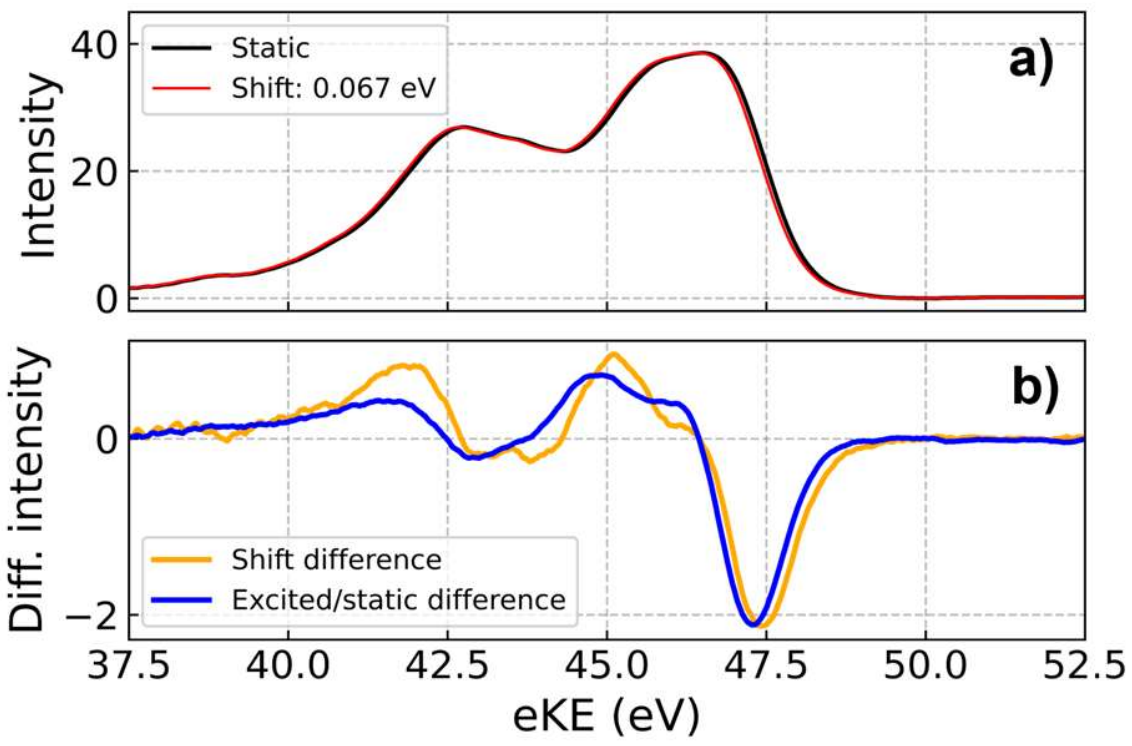

**Figure S5**. (a) Averaged experimental static spectrum of thymine over delays from 500 ps to 1 ns (black) and the same spectrum shifted by 0.067 eV (red). (b) Comparison of the averaged experimental difference spectrum (over the same 500 ps–1 ns delay range, corresponding to $T_1$ ($^3\pi\pi^*$) with the difference spectrum obtained by subtracting the experimental static spectrum from the shifted one. The close agreement shows that the triplet difference spectrum (> 6 ps) is well reproduced by a rigid energy shift of the static spectrum, consistent with reduced core-hole screening in the triplet state compared to singlet state (see main text for details).

**S5. Assignment of $I_2$ as a mixture of the $S_1$* and $T_1$ ($^3\pi\pi$*) states**

In the main text, we propose that the $I_2$ spectrum, extracted from region 3 in the difference spectrum, corresponds to a mixture of the singlet $S_1$* and triplet $T_1$ ($^3\pi\pi$*) states. To test this assignment, we extended the cascade kinetic model: $S_2 \rightarrow S_1 \rightarrow I_1$ with a relaxation branch: $S_1 \rightarrow S_1^* \rightarrow I_1$. The corresponding set of ordinary differential equations (ODEs) is:

$$\frac{dS_2}{dt} = g(t) - k_1 S_2$$

$$\frac{dS_1}{dt} = k_1 S_2 - k_2 S_1 - k_3 S_1$$

$$\frac{dI_1}{dt} = k_2 S_1 + k_4 S_1^*$$

$$\frac{dS_1^*}{dt} = k_3 S_1 - k_4 S_1^*$$

where the rate constants $k_1$ and $k_2$ correspond to $S_2$ and $S_1$ decay in the primary cascade kinetic model, while $k_3$ and $k_4$ describe the decay along the relaxation branch ($S_1 \rightarrow S_1^* \rightarrow I_1$).

We applied the same global analysis technique described in Section S3. The optimization process terminated successfully with a reduced chi-squared value ($\chi_r^2$) of 0.95 very close to 1, indicating excellent agreement between the fitted data and the experimental data. The extracted lifetimes are (92 ± 18) fs for $S_2$, (9.6 ± 2.2) ps for the $S_1 \rightarrow I_1$ relaxation, and (18.0 ± 10.5) ps for the $S_1 \rightarrow S_1^*$ relaxation. Combining two latter pathways yields an overall $S_1$ state lifetime of 6.3 ps. The intermediate $S_1$* state exhibits a lifetime of (240 ± 28) ps before relaxation to $I_1$ state, which overlaps well with the lifetime of (239 ± 29) ps extracted for $I_2$ in the four-state cascade model described in the previous section. In the main text, the $I_1$ state is assigned purely to the $T_1$ ($^3\pi\pi$*) state. Consequently, the $I_2$ spectrum can be interpreted as a mixture of the singlet $S_1$* and triplet $T_1$ spectral contributions. The IRF was estimated to have a standard deviation of (72 ± 5) fs.

Figures S6a and S6b display the fitted state-associated spectra and dynamic populations of the involved states in the kinetic model, respectively. Furthermore, Figure S6c shows the calculated difference spectra for the $S_1$ minimum and hot-$S_1$ state to facilitate direct comparison with the fitted experimental spectra. We observed that while the extracted $S_1$* spectrum closely resembles the calculated hot-$S_1$ spectrum in the high-electron kinetic energy (eKE) region (highlighted by the black dashed ellipse). This similarity breaks down in the low eKE-region (red dashed ellipse). This discrepancy prevents a definitive assignment of $S_1$* purely as a hot-$S_1$ state. The precise structural and electronic evaluation of the intermediate $S_1$* state during the 240 ps mixed-state window warrants further experimental and theoretical investigation.

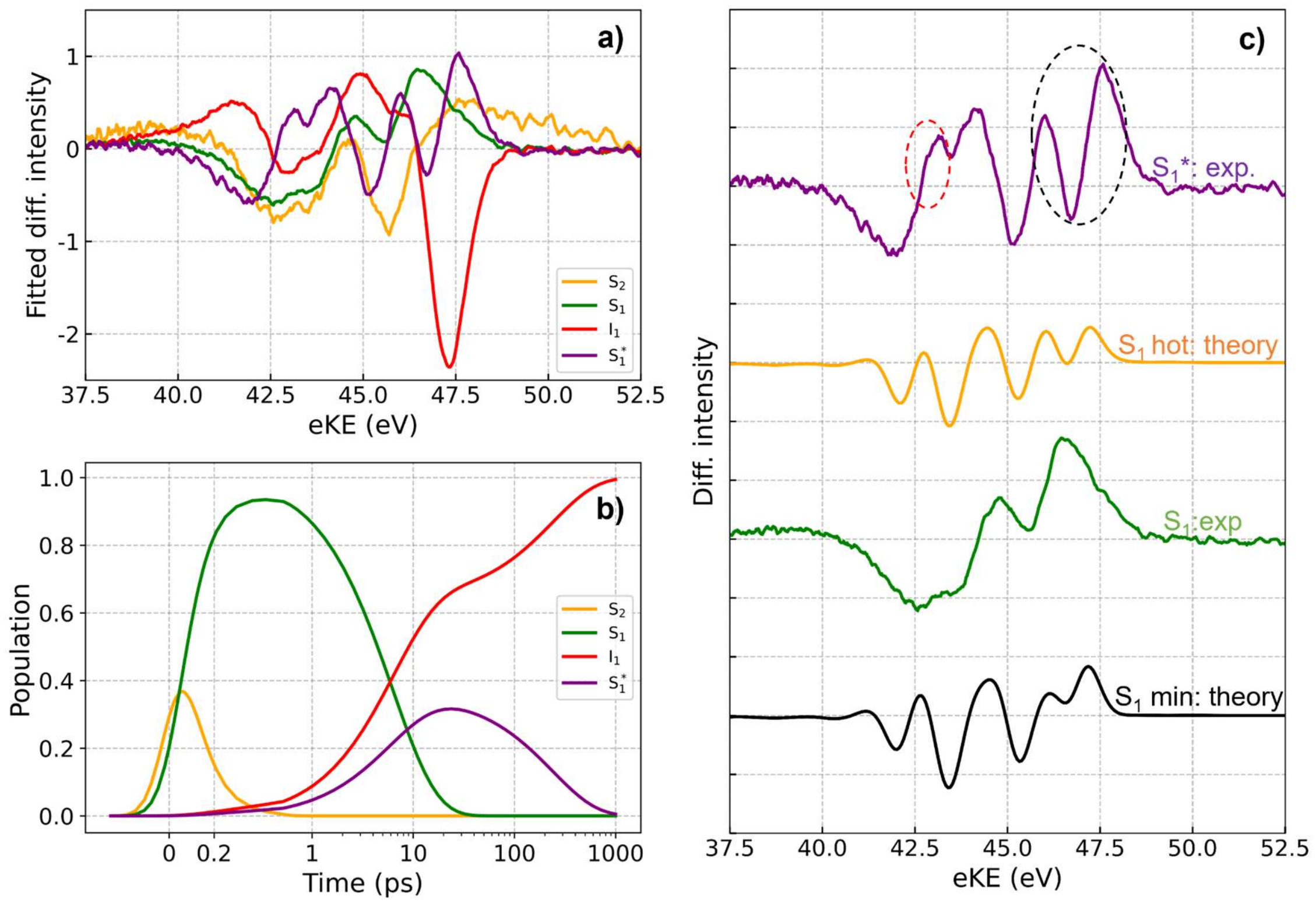

**Figure S6.** (a) Fitted state-associated spectra for the involved states in the extended kinetic model: main cascade relaxation $S_2 \rightarrow S_1 \rightarrow I_1$ with a relaxation branch $S_1 \rightarrow S_1^* \rightarrow I_1$. (b) State population dynamics extracted from the global analysis based on the extended kinetic relaxation. The fitted lifetimes for the involved states are: (92 ± 18) fs for $S_2$, (9.6 ± 2.2) ps for $S_1$ relaxation to $I_1$, and (18.0 ± 10.5) ps for $S_1$ relaxation to $S_1^*$, and (240 ± 28) ps for the intermediate state $S_1^*$. Combining the two relaxation branches from $S_1$ yields a lifetime of 6.3 ps for $S_1$ state. (c) The comparison of the fitted state-associated spectra for $S_1$ and $S_1^*$ with the calculated spectra for $S_1$ min and hot-$S_1$. The $S_1^*$ spectrum closely resembles the hot-$S_1$ spectrum in the high-eKE region (highlighted by the black dashed ellipse); however, in the low-eKE region (red dashed ellipse), this similarity degrades, which prevents a definitive assignment of $S_1^*$ state purely as a hot-$S_1$ state.

## S6. Computational details

All calculations were performed using version 24.06 of the OpenMolcas package.[7] Geometry optimizations of the ground ($S_0$) and excited states ($S_1$, $T_2$, and $T_1$), as well as conical intersection (CI) searches, were carried out with the 10 state-averaged (10SA) extended multi-state complete active space second-order perturbation theory (XMS-CASPT2) method with 14 electrons in 9 orbitals, including five π, two non-bonding (n), and two π* orbitals (denoted as the VEE version), as illustrated in Figure S7. The ANO-RCC-VDZP basis set was employed throughout. To mitigate intruder states, an imaginary shift of 0.35 Hartree was applied, while no IPEA shift was used.[8] To verify that the optimized structures correspond to true minima, frequency calculations were performed. CI searches were conducted using constrained optimizations combined with analytical evaluation of the nonadiabatic coupling vector.[9] Vertical excitation energy (VEE) calculations were subsequently performed at the 12SA-XMS-CASPT2(14e,9o) level using the same active space, which provided the most accurate reproduction of the experimental VEEs by electron energy loss spectroscopy (EELS).[10] The calculated VEEs for the $T_1$ ($^3\pi\pi^*$), $T_2$ ($^3n\pi^*$), $S_1$ ($^1n\pi^*$), and $S_2$ ($^1\pi\pi^*$) states are 3.63, 4.28, 4.50, and 4.85 eV, respectively, which are in excellent agreement with the corresponding experimental EELS values of 3.66, 4.20, 4.61, and 4.96 eV.

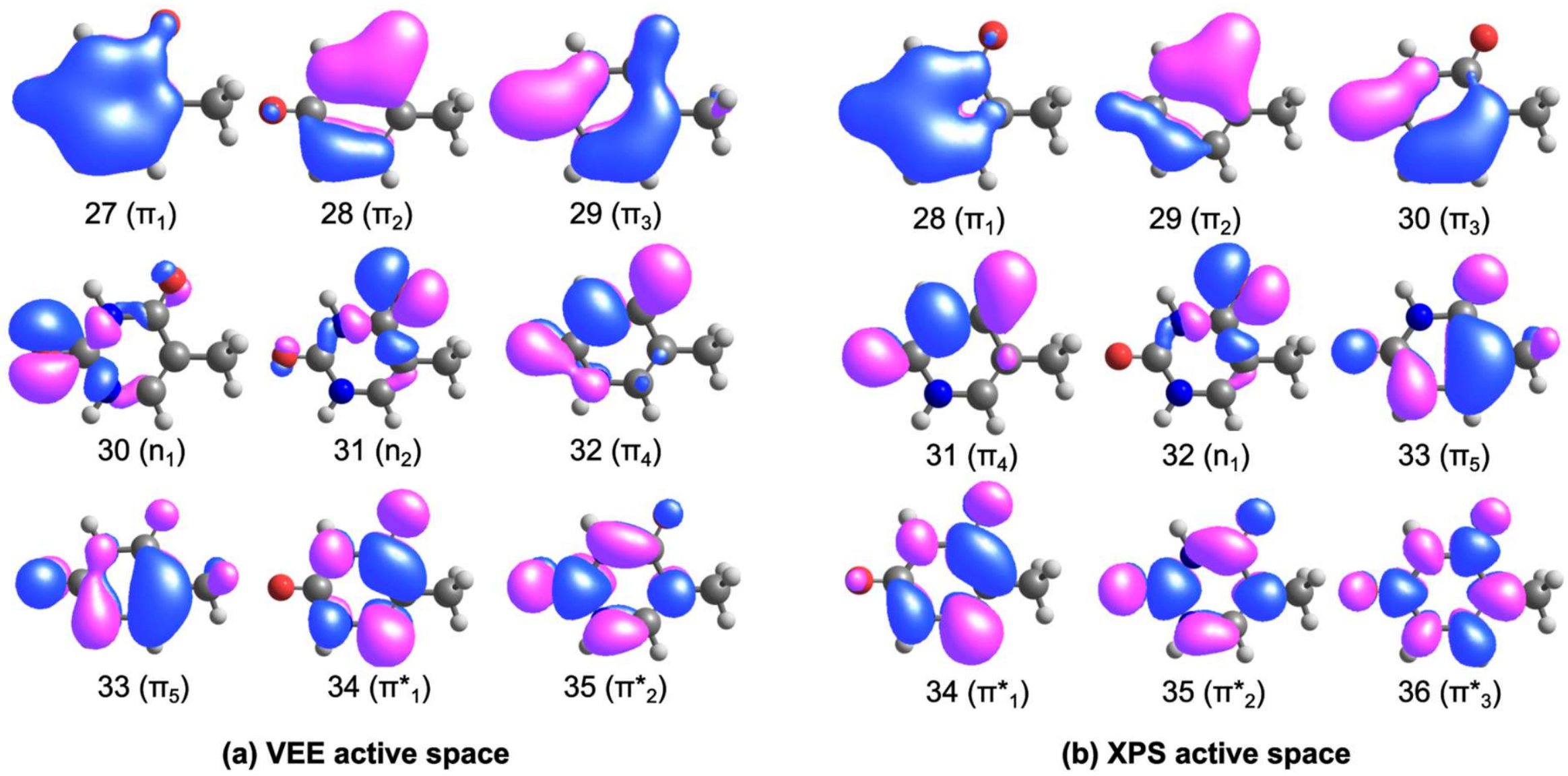

**Figure S7.** Active space for (a) 12SA-XMS-CASPT2(14e,9o) (VEE active space) and (b) 10SA-XMS-RASPT2(12e,9o) (XPS active space) calculations for singlet state at the $S_0$ minimum structure. Note that for (b), one additional core orbital is included in the active space.

XPS spectra were computed at the four optimized minima ($S_0$, $S_1$, $T_2$, and $T_1$) using the extended multi-state restricted active space second-order perturbation theory (XMS-RASPT2) method. In the restricted active space self-consistent field (RASSCF) calculations, the active space was partitioned into two subspaces: RAS1 (core orbitals) and RAS2 (valence orbitals). Core-ionized states were generated using the HEXS projection technique,[11] as implemented in OpenMolcas, which enforces a single hole in the RAS1 space. The RASSCF orbitals were separately optimized in state-averaged calculations, including 10 singlet and 10 triplet neutral states, as well as 15 doublet and 15 quartet core-ionized states. Spin-orbit coupling (SOC) was included in all XPS calculations to account for relativistic effects.

Relativistic contributions were separated into scalar relativistic effects and SOC. The scalar relativistic effects were incorporated by adding the corresponding terms of the DK2 Hamiltonian to the one-electron integrals, while SOC was treated using the spin-orbit part of the DK2 Hamiltonian.[12] This method is based on the Breit-Pauli Hamiltonian, which describes the interaction of electronic states via SOC. Finally, Dyson orbitals connecting the initial and final states were evaluated with the RASSI module in OpenMolcas, following the procedure outlined in previous literature.[13] These calculations employed XMS-RASPT2 energies together with perturbatively corrected (mixed) transition densities derived from RASSCF.

It is worth noting that the choice of active space critically influences the simulated XPS spectra. Although the VEE version of the active space successfully reproduced the experimental VEEs, it failed to accurately reproduce the ground-state ($S_0$ minimum) XPS spectra, most notably exhibiting a severe mismatch at the C9 peak position (Figure S8a). To circumvent this limitation, an alternative active space (hereafter referred to as the XPS version) was employed, which comprised five π, one n, and three π* orbitals (note that the carbon 1s core orbitals in the RAS1 space were not counted here), as illustrated in Figure S7b. As demonstrated in Figure S8b, this tailored active space successfully resolved the discrepancy, accurately capturing the C9 peak position while maintaining excellent overall agreement with all other spectral features of the experimental spectrum.

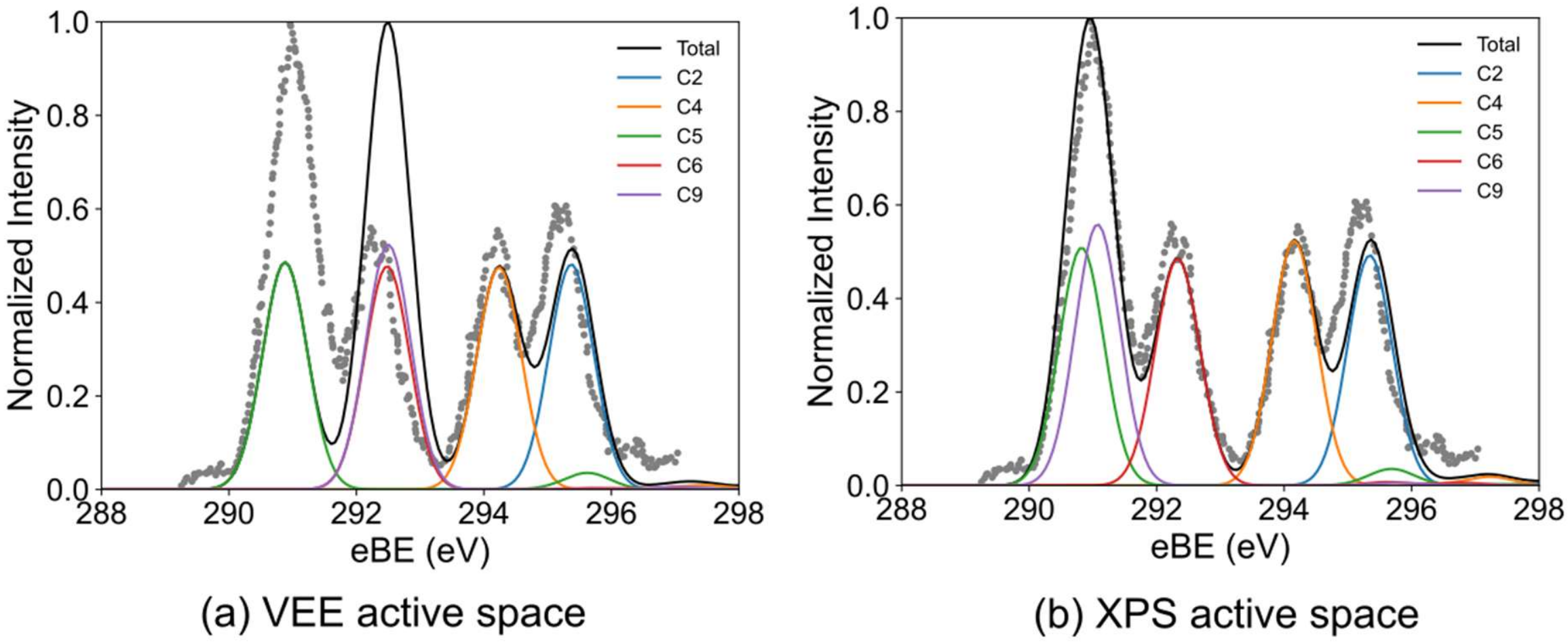

**Figure S8.** Ground state XPS spectra from two different flavors compared with the previous experimental data (gray dots).[14] To match the experimental spectrum, energy shifts (eV) of (a) −0.9 and (b) −1.1 eV of electron binding energy (eBE) were applied. All calculations utilize 0.35 of broadening factor and are computed at S0 minimum structure. Total convoluted signal is shown in black solid line while the contributions from each carbon are shown in different colors.

## S7. Details of X-ray photoelectron spectra computation

The XPS spectrum can be obtained using Dyson orbitals and their corresponding norms. A Dyson orbital is defined as the overlap between the initial *N*-electron state and the final ionized (*N*–1)-electron state (Eq. 1). By evaluating these orbitals, we extract essential information about ionization probabilities, thereby providing key insights into molecular photoionization processes.

$$\varphi_{if}^{\text{Dyson}}(x_1) = \sqrt{N} \int dx_2 \dots dx_N \Psi_f^{*(N-1)}(x_2, \dots, x_N) \Psi_i^N(x_1, x_2, \dots, x_N) \tag{1}$$

The XPS spectrum is constructed by convoluting the vertical ionization energies ($\varepsilon_0$) with the Dyson intensities (norms), using a Gaussian line-shape function (Eq. 2), as expressed in Eq. 3.

$$f_{\text{Gaussian}}(\varepsilon;\ \varepsilon_0) = \exp\left[-\frac{(\varepsilon - \varepsilon_0)^2}{2\sigma_\varepsilon^2}\right] \tag{2}$$

$$I_{\text{XPS}}(\varepsilon) = \sum_i \text{norm}_i \cdot f_{\text{Gaussian}}(\varepsilon;\ \varepsilon_{0,i}) \tag{3}$$

where $\sigma_\varepsilon$ (set to 0.35 eV) is a broadening parameter chosen to reproduce the experimental spectral resolution.

For an initial singlet state, ionization gives access exclusively to doublet final states, whereas a triplet initial state can yield both doublet and quartet final states. To properly describe SOC-induced mixing, singlet, doublet, triplet, and quartet states were computed separately and subsequently coupled through SOC interactions. Since the calculations assume equal population of the three triplet microstates, the triplet intensities were divided by three to correct for degeneracy. Dyson orbitals connecting the initial and final states were obtained with the RASSI module in OpenMolcas, following the procedure outlined in previous literature.[13] XMS-RASPT2 energies were used in combination with perturbatively corrected transition densities derived from RASSCF. To enable direct comparison with experiment, the computed spectra were uniformly shifted in energy. Specifically, the singlet spectra were shifted by 1.35 eV and the triplet spectra by 0.35 eV in eKE.

## S8. Benchmarking the choice of computational framework for XPS.

To evaluate the influence of the active spaces and basis sets on the core-level spectra, we benchmarked the ground-state XPS of thymine using three different frameworks (Figure S9). Expanding either the active space or the basis set size relative to the reference XPS version leads to noticeable deviations in the relative spectral features.

Increasing the basis set size to ANO-RCC-VTZP while maintaining the (12e,9o) active space (Figure S9a) yields an excellent absolute binding energy calibration with a 0.0 eV shift; however, it underestimates the peak position of the C6, C4, and C2 carbons in eBE. When an additional nonbonding orbital is included in the active space (Figure S9b) with the same basis set ANO-RCC-VDZP, the binding energy of the C9 peak is significantly overestimated, incorrectly appearing around 292 eV. Finally, the largest computational scheme tested, XMS-RASPT2(14e,10o) with ANO-RCC-VTZP basis set (Figure S9c), requires a +0.7-eV shift, and its predicted spectral profile deviates significantly from the experimental data.

Taken together, these benchmarks demonstrate that a careful balance between basis set size and active space is crucial for accurately reproducing the ground-state XPS spectrum. The XMS-RASPT2(12e,9o)/ANO-RCC-VDZP (XPS version) scheme provides the most faithful representation of the relative peak positions and intensities. Therefore, this computational framework was selected for ground state XPS.

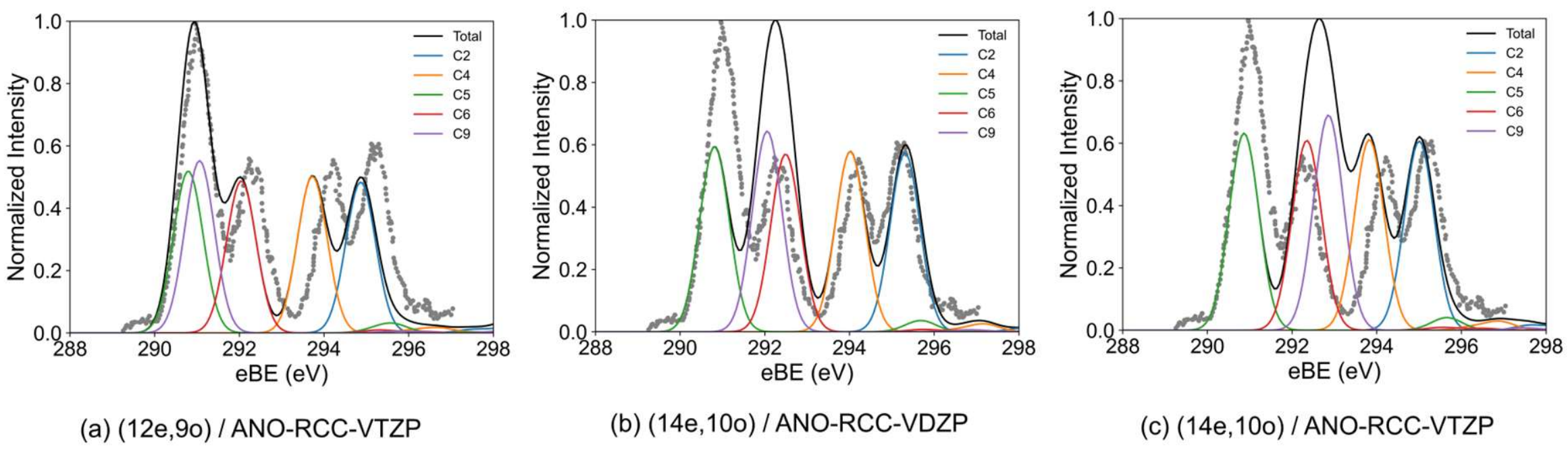

**Figure S9.** Ground state XPS spectra of thymine from three different flavors compared with the previous experimental data (gray dots).[14] To match the experimental spectrum, energy shifts (eV) of (a) 0.0, (b) -1.1 eV, and (c) +0.7 of eBE were applied. All calculations utilize 0.35 of broadening factor and are computed at $S_0$ minimum structure. Total convoluted signal is shown in black solid line while the contributions from each carbon are shown in different colors.

## S9. XPS for triplet state

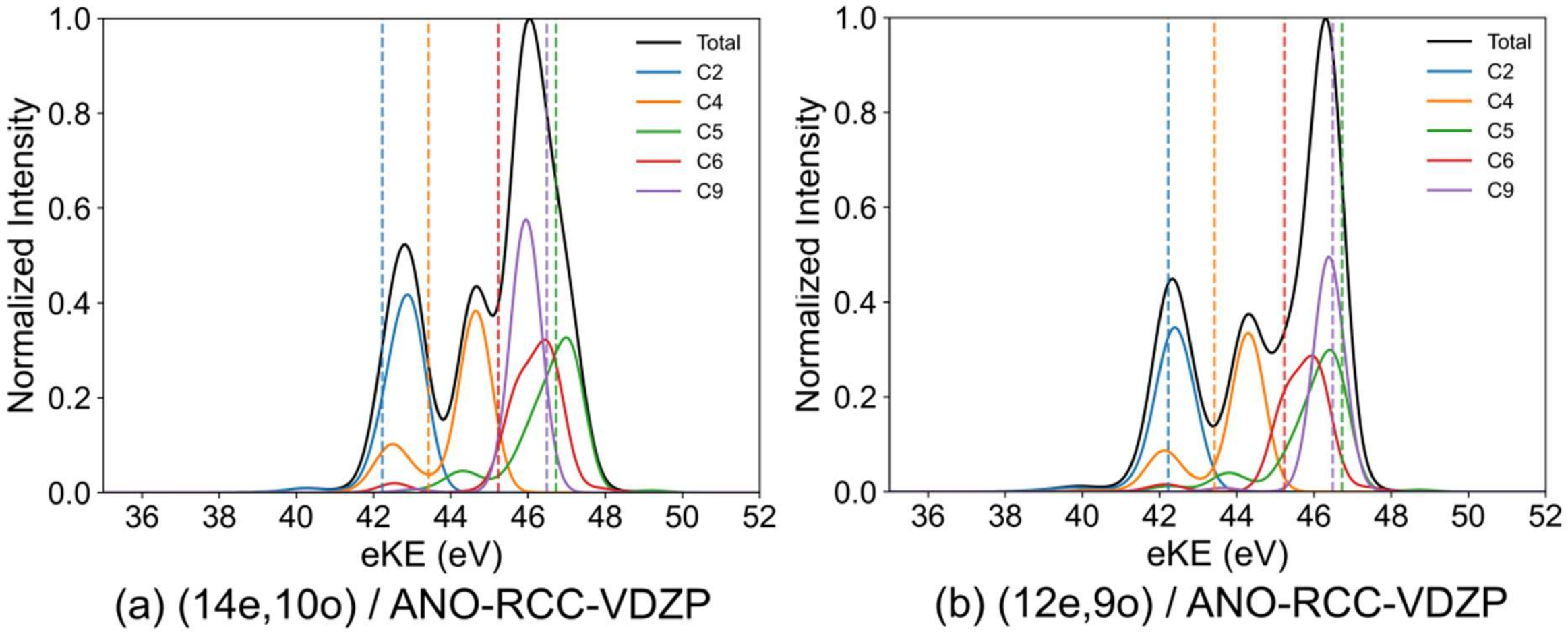

**Figure S10.** Simulated carbon 1s XPS spectra of thymine $T_1$ state using (a) (14e,10o) and (b) (12e, 9o) active space with ANO-RCC-VDZP basis set in eKE regions. All calculations utilize a broadening factor of 0.35 eV and the same energy shift of 1.35 eV as singlet spectrum. The total convoluted signal is shown as a black line, while the contributions from individual carbon centers are distinguished by different colors. Dashed vertical lines represent the maximum peak positions of the simulated $S_0$ spectrum.

Distinct active spaces and energy shifts were employed for the singlet and triplet manifolds. To enable direct comparison with the experiment, the singlet spectra were shifted by 1.35 in eKE, while the triplet spectra were shifted by 0.35 eV. Furthermore, while a (12e, 9o) active space was sufficient for the singlet state, an expanded (14e, 10o) active space was essential to correctly reproduce XPS spectra of the triplet state.

The necessity of this expanded active space for the triplet state ($T_1$ as an example) is illustrated in Figure S10. A crucial spectral signature of the $I_1$ and $I_2$ states is the distinct shift of the C9 (purple line) peak. When the $T_1$ spectrum is simulated using the proper (14e, 10o) active space (Figure S10a), the C9 peak exhibits a pronounced shift to a lower eKE, clearly separating it from the adjacent C5 and C6 peaks. In contrast, when the smaller (12e, 9o) active space is applied to the $T_1$ state (Figure S10b), this characteristic shift is entirely missing, causing the C9 signal to incorrectly overlap with the C5 peak. Consequently, the (14e, 10o) active space is essential to capture the unique signature C9 shift inherent to the triplet state.

## S10. XPS for hot $S_0$ and $T_1$ states

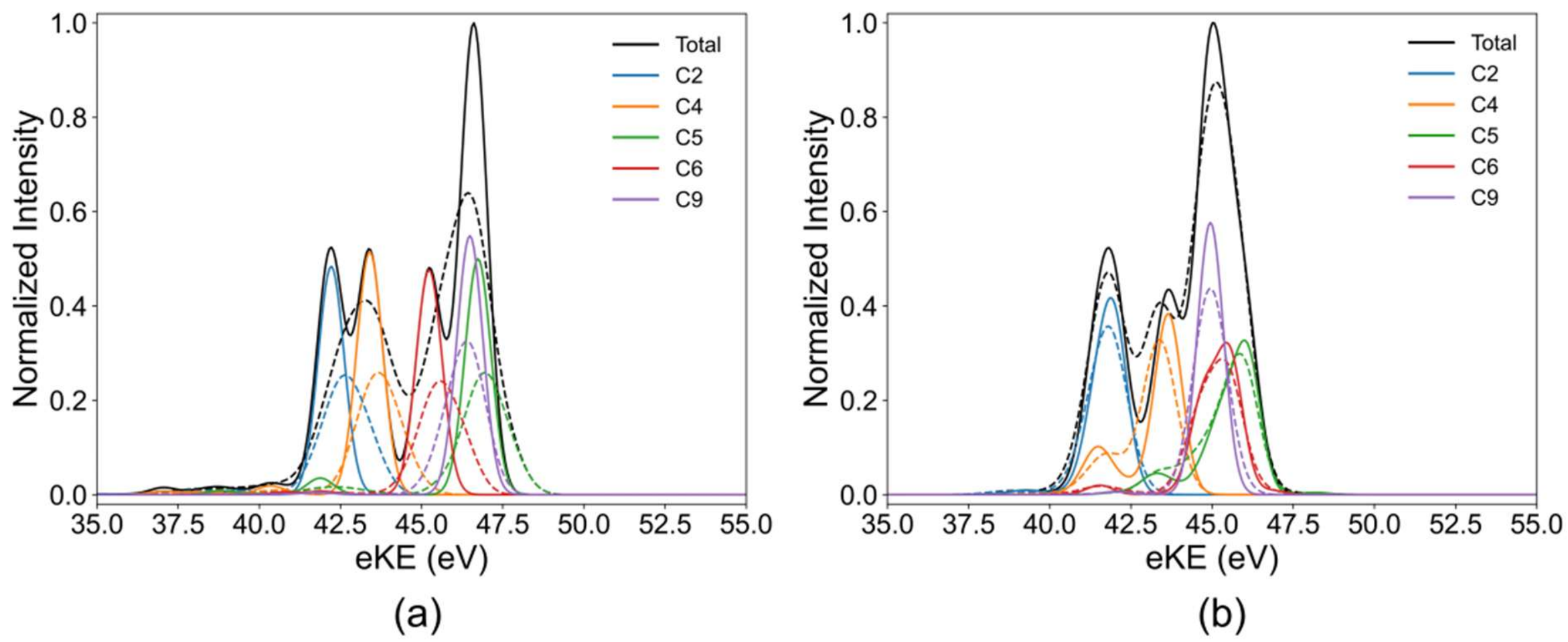

**Figure S11.** Simulated carbon 1s XPS spectra of thymine from (a) hot ground state ($S_0$) and (b) hot triplet state ($T_1$) sampled during ADMP molecular dynamics simulations shown in eKE regions. All calculations utilize a broadening factor of 0.35 eV. The total convoluted signal is shown as a black line, while the contributions from individual carbon centers are distinguished by different colors. Solid and dashed lines represent the spectra obtained from the equilibrium and averaged hot trajectories, respectively.

To assess the possible contribution of hot ground state ($S_0$) and hot triplet state ($T_1$) to the tr-XPS signal, we performed carbon 1s ground-state XPS calculations using structures sampled from atom-centered density matrix propagation (ADMP) simulations[15] with the Gaussian16 program.[16] The ADMP molecular dynamics at the B3LYP/6-31G* level of theory were propagated with an initial nuclear kinetic energy of 5 eV for the hot ground state, corresponding to the experimental vertical excitation energy of the $\pi\pi^*$ state, whereas 1.84 eV was applied for the hot $T_1$ state, corresponding to the energy difference between the vertical excitation energy (5 eV) and the optimized $T_1$ minimum structure (3.16 eV).

From these trajectories, 300 geometries were randomly selected to compute the respective hot-state XPS spectra and compared in Figure S11. The results show that for both the hot $S_0$ and hot $T_1$ manifolds, the positions of the individual carbon peaks remain nearly unchanged in terms of their central eKE region. Instead, the high internal vibrational energy primarily manifests as a significant broadening of the spectral features due to thermal nuclear motion. Crucially, neither the hot ground state nor the hot triplet state calculations exhibit the distinct spectral shifts required to explain the negative features observed in region R3 and R4. This quantitative analysis indicates that the experimentally observed negative band at 47.5 eV cannot be attributed to hot ground state or hot triplet state contributions.

## S11. Ground-state intermediate

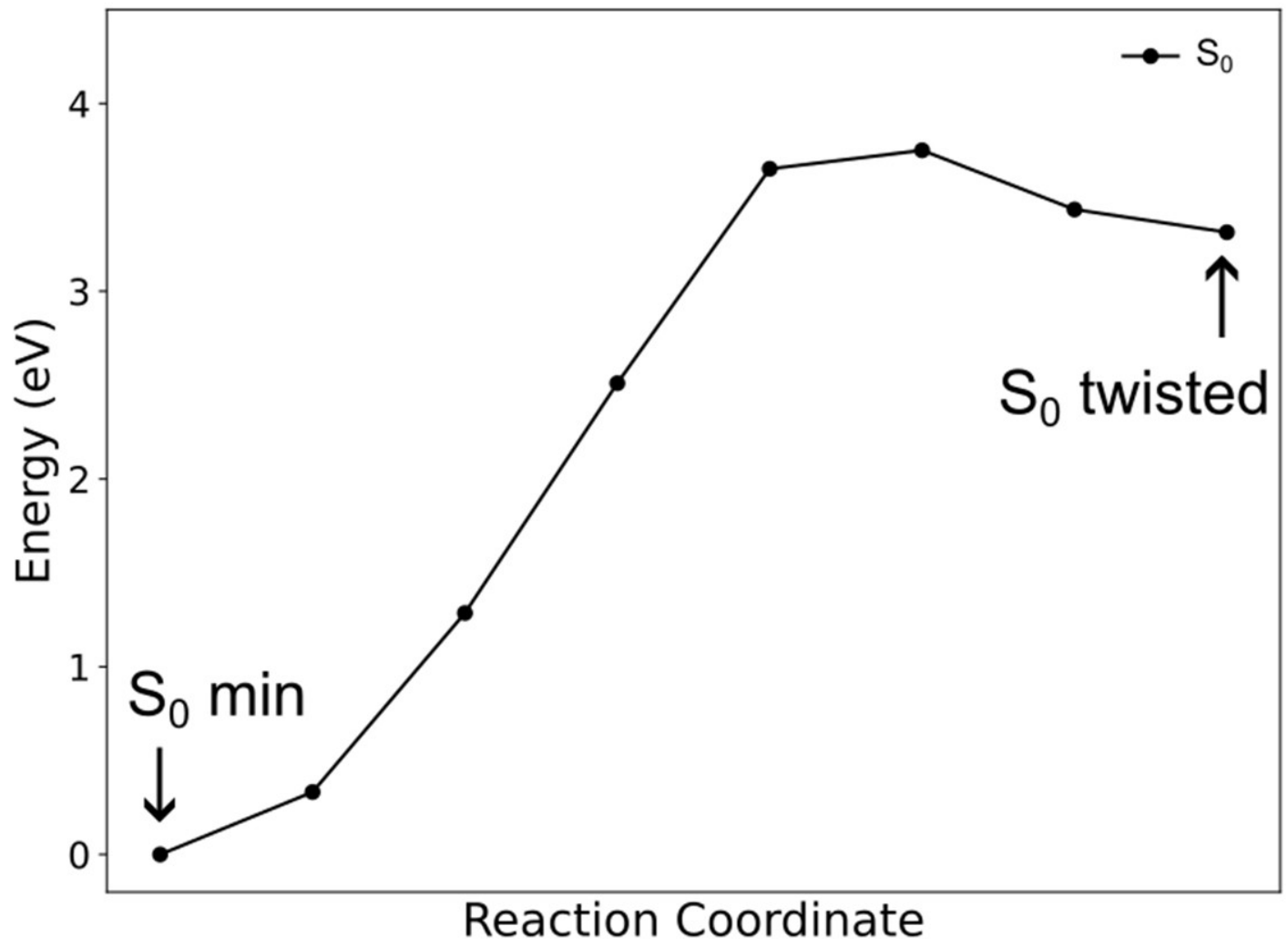

**Figure S12.** Potential energy curves (PECs) of thymine along the reaction coordinates from the $S_0$ minimum to $S_0$ twisted intermediate structure. The energy (in eV) of the $S_0$ state is plotted in black.

According to our previous study,[17] it was proposed that uracil, upon photoexcitation, rapidly traverses the $\pi\pi^*/S_0$ conical intersection, partially yielding an $S_0$ intermediate characterized by a twisted C5=C6 bond rather than returning directly to the planar ground-state structure. Recent experimental studies have further confirmed that this twisted conformation is kinetically stable and exhibits a relatively long lifetime.[18–20] Although thymine does not strictly follow the same direct $\pi\pi^*/S_0$ deactivation pathway, minimum energy path (MEP) calculations indicate that the $T_1$ ($\pi\pi^*$)/$S_0$ conical intersection geometry closely resembles that of the $S_1$ ($\pi\pi^*$)/$S_0$ intersection (Figure S17a). This structural similarity suggests the plausible formation of an analogous $S_0$ intermediate via the $T_1 \rightarrow S_0$ decay channel.

However, the MEP connecting the $S_0$ minimum and the twisted $S_0$ intermediate shows a barrier of around 0.45 eV (Figure S12) from the twisted conformation back to the planar ground state. Given the substantial excess vibrational energy deposited into the ground state upon nonradiative deactivation, such a moderate barrier is not sufficiently high to sustain a long-lived population. Consequently, the twisted $S_0$ intermediate cannot account for the long lifetimes of the species experimentally observed in region R4 in the experimental difference spectrum, allowing us to rule it out from its assignment.

**S12. Origin of different phase along ring-breathing mode**

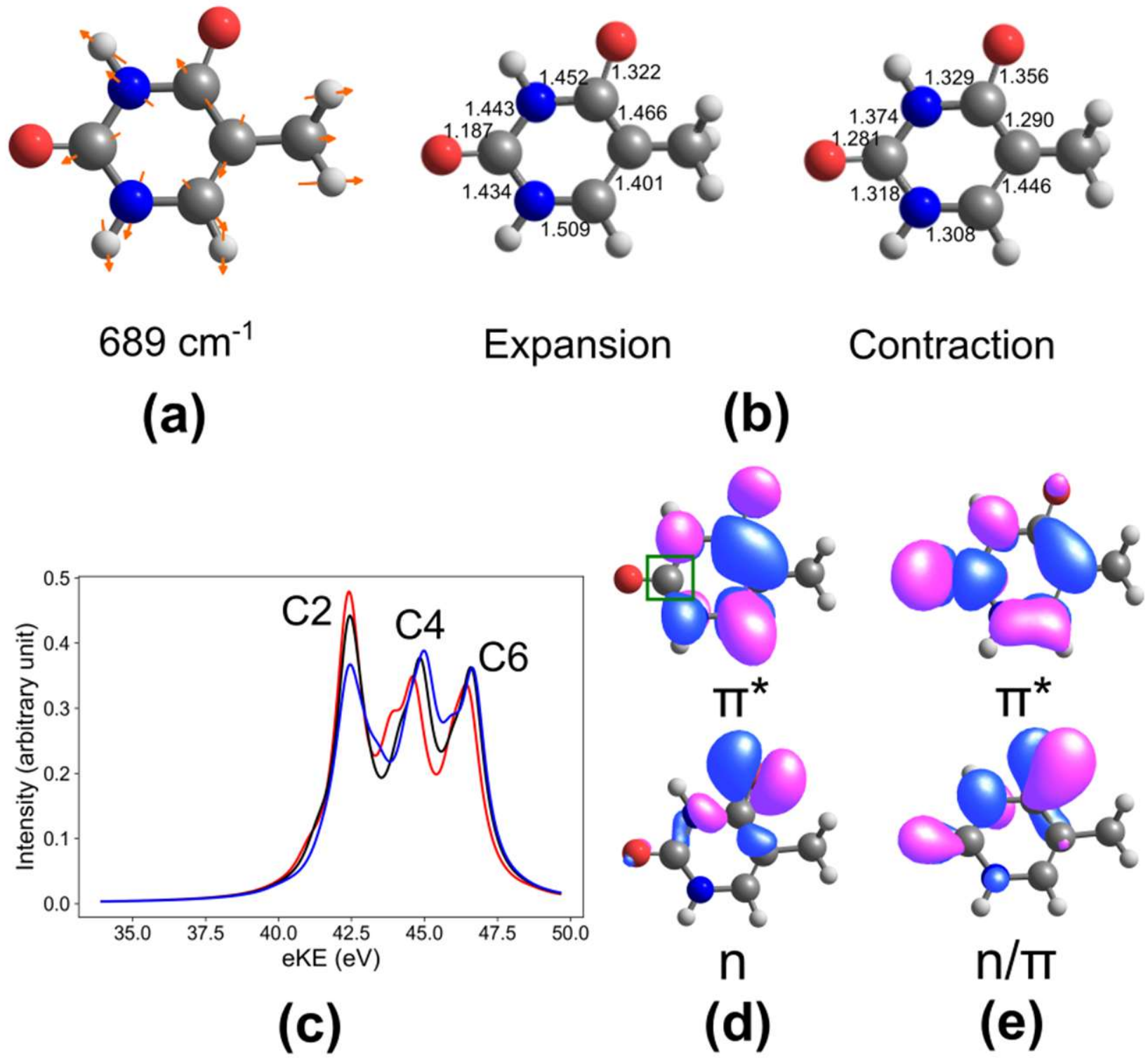

**Figure S13.** (a) Normal mode displacement vector of the ring-breathing mode at the $S_1$ minimum (689 cm$^{-1}$, shifted from 730 cm$^{-1}$ in the $S_0$ state). (b) Geometrical structures corresponding to the ring expansion and contraction along the 689 cm$^{-1}$ mode with bond lengths shown in Å. (c) Simulated photoelectron spectra at $S_1$ minimum (black), expanded (red), and contracted (blue) geometries. Only C2, C4, and C6 peaks are shown. (d,e) Molecular orbitals (MOs) of core-ionized cationic state (doublet state shown only) corresponding to photoionization at (d) the C4 and C6 sites and (e) the C2 site, illustrating the distinct MO configuration. The green box indicates zero electron density at the C2 atom.

The site-specific spectral modulation finds its electronic origin in the distinct orbital response of the core-ionized cationic states to nuclear deformation. As illustrated in Figure S13d, photoionization at the C4 and C6 sites leaves the valence n and π* orbitals largely unchanged compared to the neutral state (See Figure S7a MO34); the π conjugation remains largely along the C4=C5-C6=O8 region, leaving a nodal plane across the C2 atom. Conversely, photoionization at the C2 site (Figure S13e) induces a spatial redistribution of the valence π* orbital, shifting the conjugation prominently toward the C2=O7 region (Figure S13e). Due to these contrasting local charge distributions, the valence orbitals undergo opposing responses along the ring-breathing vibrational mode, which drives the out-of-phase (π phase shift) intensity modulation observed in the XPS spectrum (Figure 3b).

## S13. Normal mode projection

To quantitatively monitor the evolution of specific vibrational modes during the nonadiabatic relaxation, the excited-state trajectories from the previous literature were projected onto the normal modes of the ground-state equilibrium structure ($S_0$ minimum).[21] For each trajectory frame at time t, the Cartesian displacement vector ΔR(t) from the reference $S_0$ geometry ($R_{min}$ (t=0)) was computed after performing a mass-weighted Kabsch alignment to remove overall translation and rotation:

$$\Delta R(t) = R_{aligned}(t) - R_{min}(t = 0)$$

To generate the mass-weighted normal coordinates ($q_k$), the Cartesian displacement vector was scaled by the square root of the atomic masses ($M^{1/2}$) and subsequently projected onto the orthonormal eigenvectors ($L_k$) of the internal coordinate mass-weighted Cartesian Hessian matrix.

$$q_k(t) = {L_k}^T \cdot M^{\frac{1}{2}} \cdot \Delta R(t)$$

Because the Hessian eigenvectors $L_k$ are dimensionless and normalized to unity, the resulting projection amplitude explicitly retains the $amu^{1/2}\cdot$Å dimensional unit. This projection analysis allows for a direct mapping of the nuclear wavepacket dynamics onto the specific normal coordinates of interest.

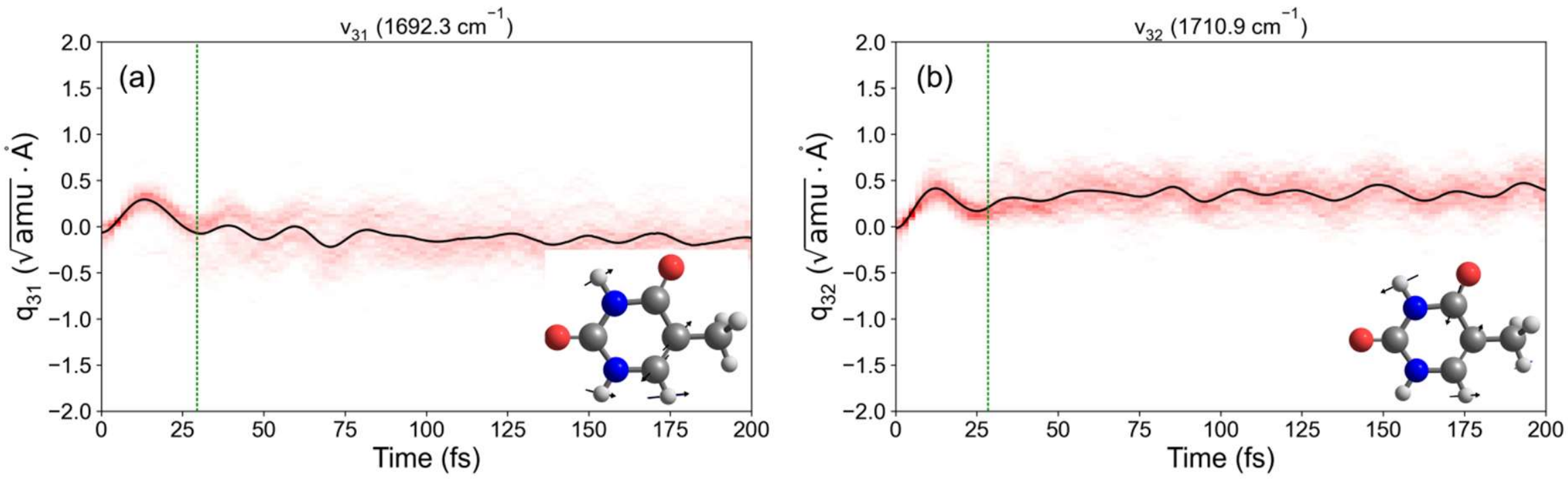

**Figure S14.** Time-evolution of the mass-weighted normal coordinates (a) $q_{31}$ (C5=C6 stretch) and (b) $q_{32}$ (C4=O8 stretch) of thymine. Excited-state trajectories are projected onto the normal modes of the $S_0$ minimum. The red 2D histogram displays the overall trajectory distribution, and the black line indicates the ensemble average. The vertical green dashed line (~30 fs) marks the completion of the first period of the BLA motion and the subsequent abrupt attenuation of the vibrational coherence. The insets show the corresponding ground-state normal mode displacement vectors.

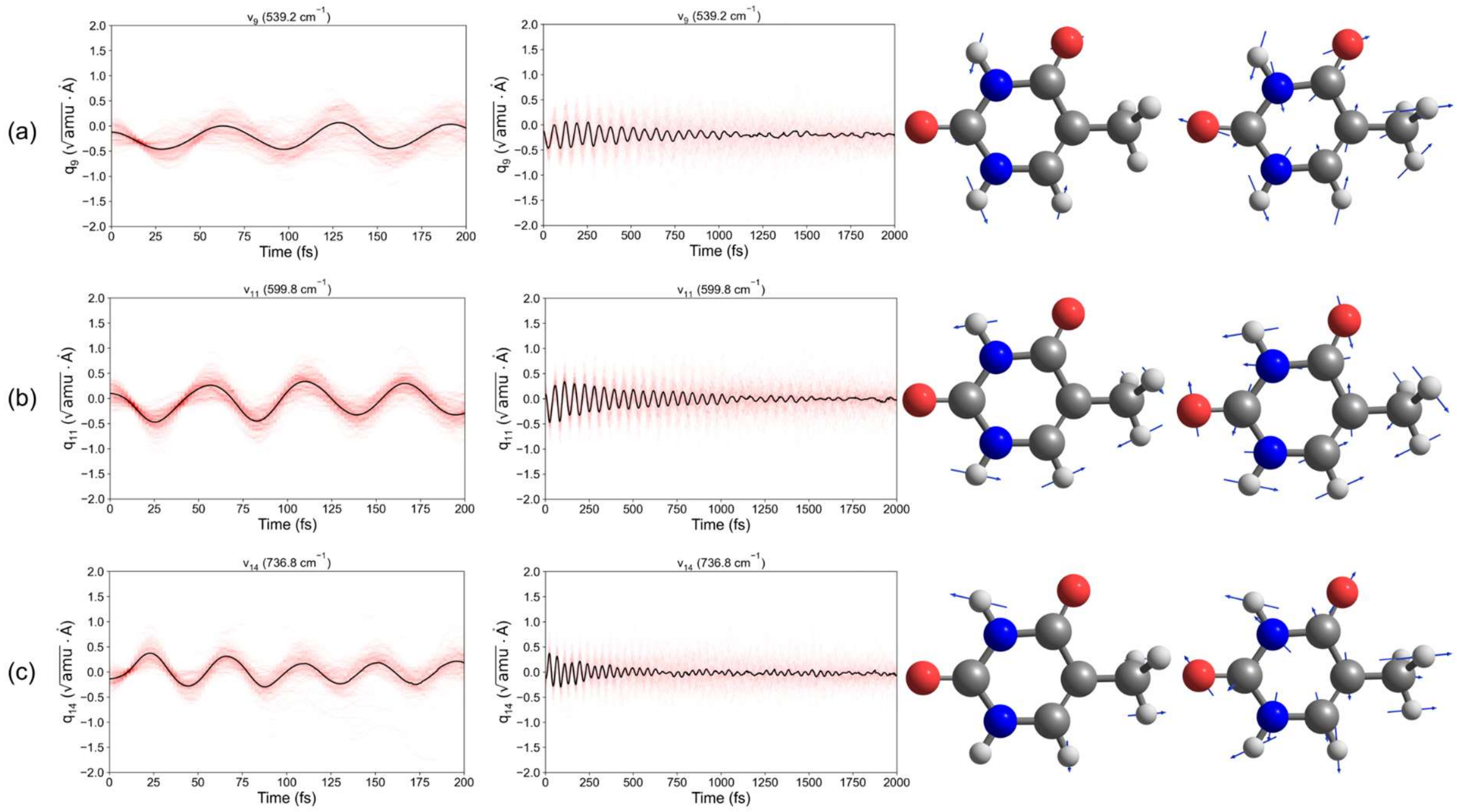

**Figure S15.** Same analysis as Figure S14 for the selected coherent low-frequency mass-weighted normal coordinates (a) $q_9$, (b) $q_{11}$, and (c) $q_{14}$ of thymine. The first column presents the short-time dynamics up to 200 fs, while the second column extends the same analysis up to 2 ps. The third and fourth columns show the mass-weighted and unscaled normal mode displacement vectors, respectively.

### S14. Normal mode scan

Potential energy curves along the normal modes were constructed by systematically displacing the $S_1$ minimum geometry along dimensionless normal coordinates $Q_k$. For a given normal mode k, the Cartesian displacement vector was generated from the dimensionless normal coordinate according to

$$\Delta R_k = Q_k \cdot e_k \cdot \sqrt{\hbar/(m_k \cdot \omega_k)}$$

where $Q_k$ is the dimensionless normal coordinate, ħ is a reduced Planck constant, $e_k$ is the mass-weighted normal mode vector, $m_k$ is the reduced mass of mode k, and $\omega_k$ is the corresponding vibrational frequency. Geometries were generated across Q = −8.0 to +8.0 with a step size of ΔQ = 0.5 to trace the electronic energy profiles of the singlet ($S_0$, $S_1$, $S_2$) and triplet ($T_1$, $T_2$, $T_3$) states.

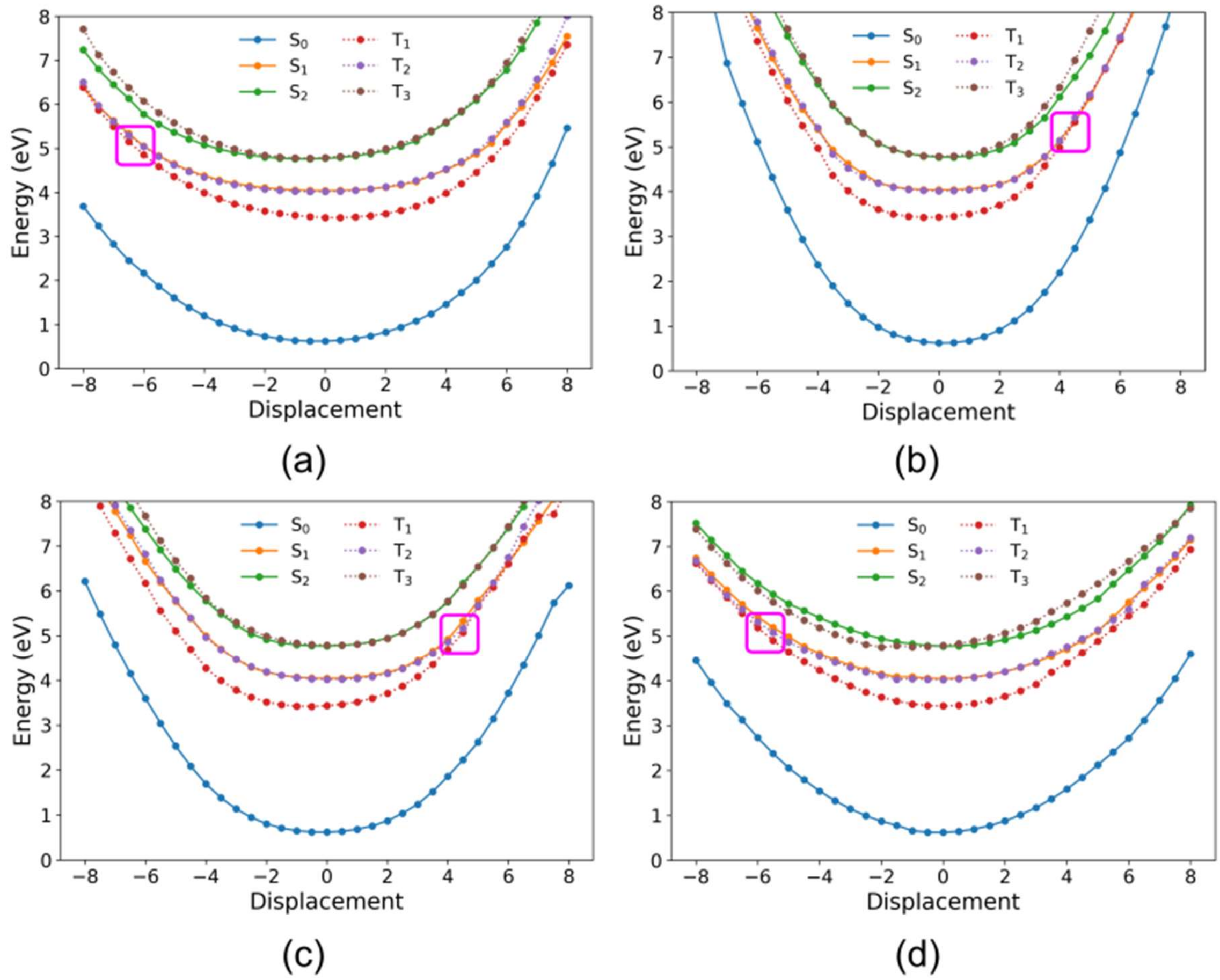

**Figure S16.** Potential energy curves (PECs) of thymine computed along representative dimensionless normal coordinates (Q) for out-of-plane normal modes (a) 231 $cm^{-1}$, (b) 262 $cm^{-1}$, (c) 350 $cm^{-1}$, and (d) 546 $cm^{-1}$. Magenta boxes indicate possible three state crossing regions.

**S15. Three state crossing point optimization**

To locate the three-state crossing point among the lowest excited singlet state ($S_1$) and the two lowest triplet states ($T_1$ and $T_2$), a home-written Python script interfaced with the OpenMolcas program package was employed. The electronic structures, including total energies and gradients of each state, were evaluated at the 10SA-XMS-CASPT2(14e,9o) level of theory. The multi-state intersection optimization was driven by the penalty function method,[22] which systematically minimizes the average energy of the target states while driving their energy gaps to zero. The multi-state penalty objective function optimized in this study is defined as follows:

$$O_{penalty} = E_{avg} + \sigma \sum_{(a,b)} \frac{\{E_a - E_b)^2}{|E_a - E_b| + \alpha}$$

Where $E_{avg}$ represents the mean of the total XMS-CASPT2 energies of the three target states ($S_1$, $T_2$, and $T_1$).

The parameter σ acts as the penalty weight that controls the coupling strength between the energy minimization and the gap closure. The parameter α is a smoothing factor introduced to damp the singularities in the energy gradients near the crossing region, preventing state-swapping numerical instabilities. The coupled analytical gradients of the objective function ($dO_{penalty}$) with respect to the nuclear coordinates were evaluated at each step to update the molecular geometry via the steepest descent method:

$$R_{n+1} = R_n - \gamma \cdot dO_{penalty}$$

where R represents the nuclear coordinate vector, n is the iteration index, and γ is the optimization step size.

## S16. Minimum energy potential energy curves (PECs)

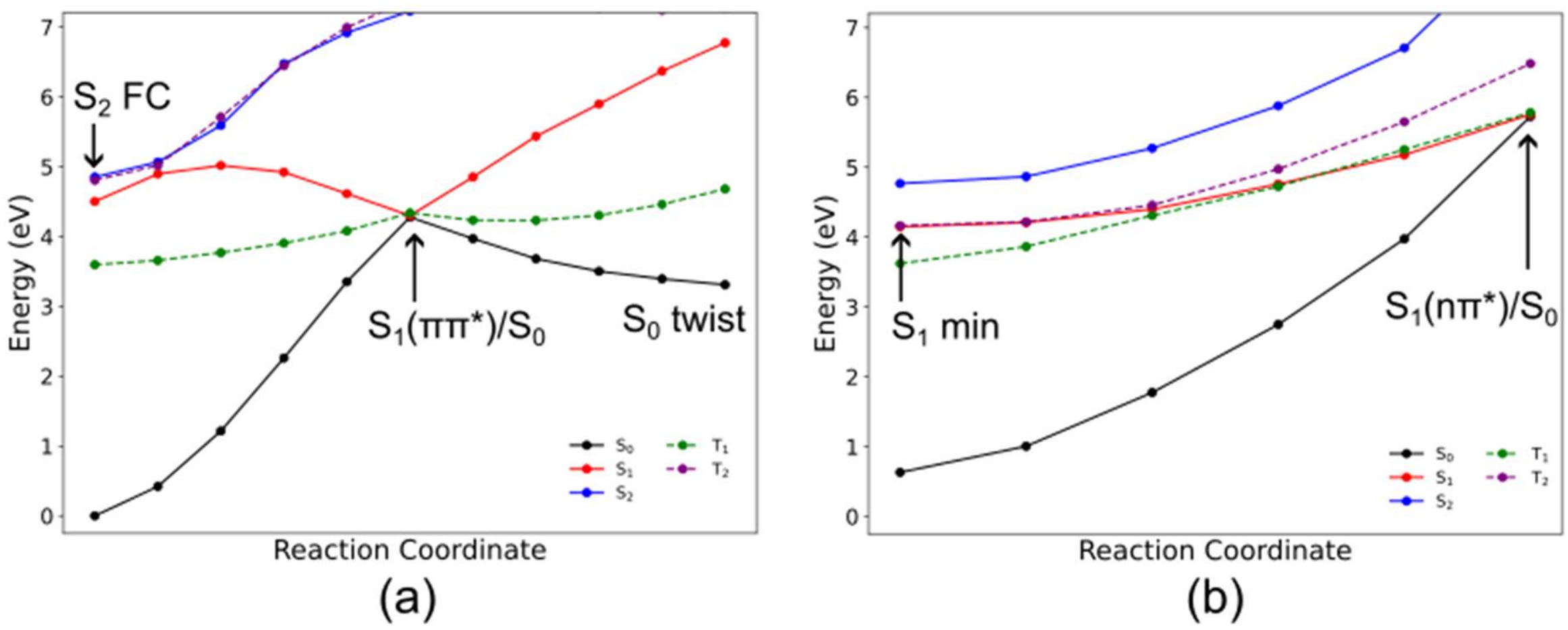

**Figure S17.** Potential energy curves (PECs) of thymine along the reaction coordinates (a) from the $S_2$ Franck-Condon (FC) towards the $S_1(\pi\pi^*)/S_0$ conical intersection and the subsequent $S_0$ twist intermediate; and (b) from the $S_1$ minimum towards the $S_1(n\pi^*)/S_0$ conical intersection. The energies (in eV) of $S_2$, $S_1$, $T_2$, and $T_1$ are respectively plotted in blue, red, purple, and green, with singlet and triplet states represented by solid and dashed lines, respectively.

### S17. Hyperconjugation of methyl group

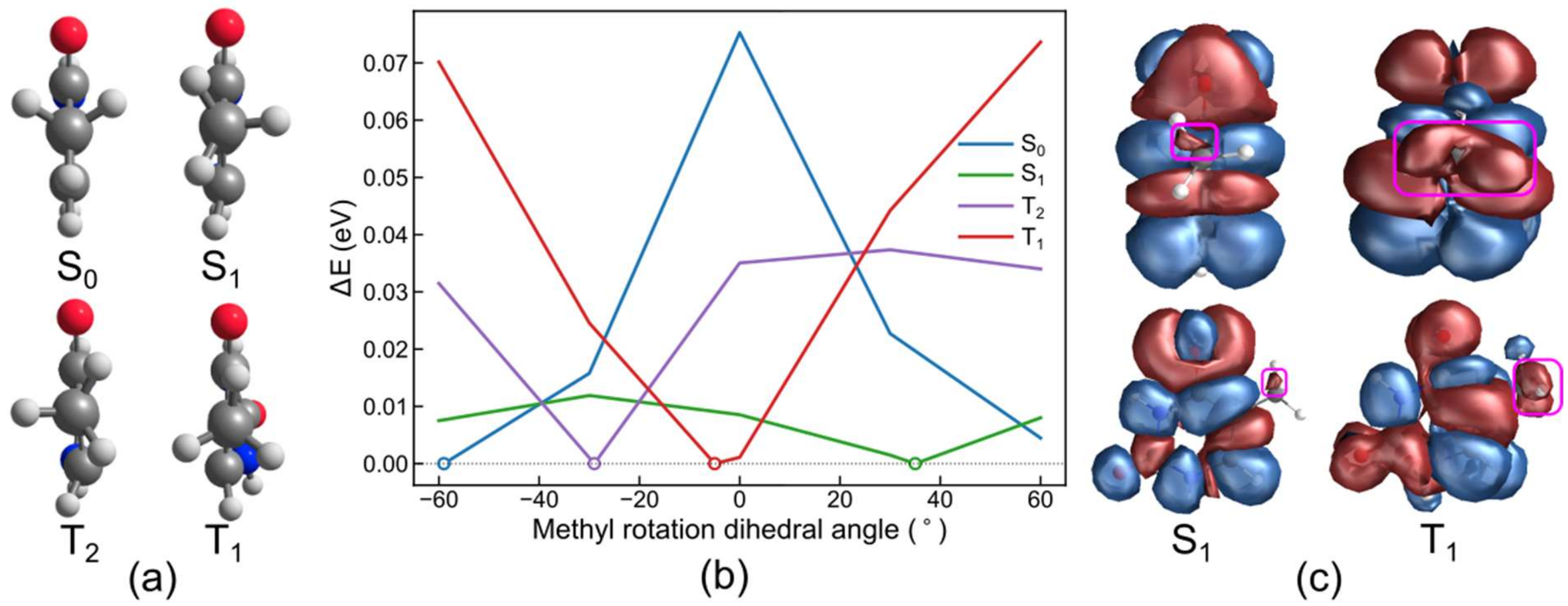

**Figure S18.** (a) Side-view of the optimized minimum energy structures of the $S_0$, $S_1$, $T_2$, and $T_1$ states, highlighting the different methyl-group orientations. (b) Potential energy profiles along the methyl rotation coordinate for the $S_0$, $S_1$, $T_2$, and $T_1$ states. The methyl rotation minimum energy dihedral angles coordinate is indicated as open circles, correspondingly at -59°, 35°, -29°, and -5° for $S_0$, $S_1$, $T_2$, and $T_1$. Energies are given in eVs relative to the respective minimum. (c) Attachment/detachment density analysis for the $S_1$-$S_0$ transition at the $S_1$ minimum geometry and the $T_1$–$S_0$ transition at the $T_1$ minimum geometry. Top row is a side view like in (a), bottom row presents a top view onto the ring. Red and blue isosurfaces denote detachment and attachment densities, respectively. The highlighted regions with pink box indicate the methyl-group contribution to the electron density redistribution through hyperconjugation.

The optimized structures show that the methyl rotation is strongly dependent on the electronic state (Figure S18a). At the $S_0$ minimum, the methyl group adopts a dihedral angle of -59°, where two C-H bonds are approximately aligned with the conjugated π-framework. This orientation allows effective hyperconjugative interaction between the methyl $\sigma_{C\text{-}H}$ orbitals and the molecular π-conjugation. In contrast, the $S_1$ and $T_2$ minima, which have predominant nπ* characters are located at 35° and -29°, respectively. These geometries are rotated by approximately 30° relative to the optimal hyperconjugative arrangement, reducing the alignment of the C-H bonds with the π-system and thereby weakening the hyperconjugative interaction.

Interestingly, the $T_1$ minimum is found at a methyl rotation angle of -5°. In this geometry, two C-H bonds again become favorably oriented with respect to the π-framework, but with a distinct torsional phase compared to the $S_0$ minimum. This indicates that hyperconjugation is restored in the $T_1$ state, although through a different methyl alignment from that in the ground state. The methyl-rotation scans (Figure S18b) further support this state-specific behavior, showing that each electronic state has a distinct minimum along the torsional coordinate.

The attachment/detachment density analysis (Figure S18c) also reveals different density redistribution patterns for the $S_1$-$S_0$ and $T_1$-$S_0$ transitions near the methyl group. The methyl group in the $T_1$ state contributes more strongly to the electronic redistribution through hyperconjugation than in the $S_1$ state, supporting the interpretation that the C9 shift at the $T_1$ minimum originates from enhanced hyperconjugation of the methyl group.

**S18. Cartesian coordinates optimized at 10SA-XMS-CASPT2(14e,9o)/ANO-RCC-VDZP**

**$S_0$ min**

C -6.28418536 3.37414012 -0.28774348

C -5.04828641 2.65242768 0.01510114

C -5.13060013 1.32112579 0.28530174

C -7.55077611 1.22883753 -0.00039981

C -3.76623022 3.42610027 0.00655490

N -6.33077022 0.63945791 0.28480565

N -7.44443530 2.58550476 -0.26290951

O -8.61191982 0.60981190 -0.01668389

O -6.34773275 4.57785686 -0.54917106

H -6.36915525 -0.35175380 0.47802577

H -8.31739885 3.05969343 -0.47266196

H -3.59576555 3.87803406 -0.97598328

H -2.91815376 2.78036190 0.25089112

H -3.80718600 4.24625945 0.73060462

H -4.25801430 0.72011414 0.51599509

**$CI_{21}$ ($^1\pi\pi^*$/$^1n\pi^*$)**

C -6.27005825 3.41426474 0.00064518

C -5.06711366 2.69475935 0.00016011

C -5.08213743 1.22009894 -0.00062357

C -7.53933900 1.29094708 0.00010297

C -3.78021435 3.42290290 0.00015383

N -6.28864531 0.64445321 -0.00105007

N -7.47580765 2.63270312 0.00127748

O -8.53678276 0.60941036 -0.00001615

O -6.40400869 4.64048933 0.00056442

H -6.38400217 -0.35723043 -0.00185482

H -8.33889427 3.14147519 0.00217709

H -3.69313159 4.07485812 -0.87098890

H -2.93127227 2.74499279 0.00560707

H -3.69854309 4.08275467 0.86570020

H -4.21080461 0.59814963 -0.00185695

**$S_1$ ($^1n\pi^*$) min**

C -6.25242972 3.30669565 -0.20142310

C -5.05017673 2.66950342 0.00229346

C -5.08396647 1.26141669 0.15878358

C -7.55006939 1.22508931 -0.00036836

C -3.76331552 3.44632343 -0.00708181

N -6.33570702 0.62093080 0.22904635

N -7.45049990 2.60306220 -0.25494338

O -8.63659308 0.64327251 0.01504117

O -6.41740490 4.61784529 -0.41083824

H -6.39248538 -0.38007286 0.35846663

H -8.32830372 3.09143224 -0.37692880

H -3.37055055 3.57193993 -1.02265346

H -3.00119591 2.93692725 0.58987494

H -3.91311784 4.44591013 0.41688973

H -4.23479387 0.68769603 0.50556928

**$CI_{10}$ ($^1n\pi^*/S_0$)**

C -6.194436 3.236451 -0.352158

C -5.027423 2.636421 -0.135418

C -5.074253 1.159043 -0.090692

C -7.527089 1.355144 0.347254

C -3.861803 3.396340 0.359547

N -6.339611 0.702784 0.405961

N -7.471335 2.644044 -0.238931

O -8.604392 0.883088 0.702868

O -6.738374 4.483995 -0.004267

H -6.424769 -0.232978 0.777307

H -8.147588 2.797069 -0.976172

H -3.029720 3.500552 -0.374327

H -3.426559 2.925457 1.238163

H -4.136549 4.425965 0.636022

H -4.634339 0.519750 -0.854507

**$CI_{10}$ ($^1\pi\pi^*/S_0$)**

C -6.14770883 3.56055506 -0.32782093

C -4.89051684 2.83002907 -0.20417191

C -5.02102603 1.77551408 0.73540303

C -7.29193181 1.38166000 0.18462081

C -4.67254726 2.19950186 -1.61460695

N -6.02244181 0.90314692 0.65079384

N -7.29070987 2.69568004 -0.20647603

O -8.27269179 0.65738899 0.25148789

O -6.38332692 4.73886904 -0.57844896

H -6.10387983 0.03493489 1.17638479

H -8.20999688 3.09121903 -0.38133397

H -5.54607823 1.66446490 -2.02344207

H -3.82339826 1.50748186 -1.58943303

H -4.41769722 3.01314791 -2.30099889

H -4.37484441 1.70858635 1.61472235

**$S_1$ ($^1n\pi^*$)/$T_2$ ($^3n\pi^*$)/$T_1$ ($^3\pi\pi^*$) three state CI**

C -6.24845620 3.25840998 -0.16822268

C -5.02333988 2.61751544 -0.00312142

C -5.08450615 1.31115950 0.19484178

C -7.54633610 1.22758877 -0.02628822

C -3.76093647 3.45135064 -0.00978615

N -6.33706150 0.61125200 0.21729238

N -7.44422664 2.62962283 -0.26457002

O -8.62420785 0.62113451 0.01917072

O -6.48642851 4.61494315 -0.45169954

H -6.39357022 -0.37726308 0.35968014

H -8.33632538 3.09232359 -0.35383686

H -3.37251074 3.57357437 -1.02543099

H -3.00076478 2.94310293 0.59125109

H -3.90013400 4.44079965 0.42193152

H -4.23634412 0.69332898 0.49481864

**$T_2$ ($^3n\pi^*$) min**

C -6.24628004 3.28946234 -0.34583277

C -5.04171458 2.66125664 -0.03209984

C -5.06909647 1.22028544 0.21742623

C -7.54723177 1.21948005 -0.01051014

C -3.76767946 3.45139722 -0.00559377

N -6.35129651 0.65210523 0.38665749

N -7.42198998 2.56801061 -0.38377162

O -8.62870568 0.64021682 -0.02091710

O -6.38217279 4.58924949 -0.63003516

H -6.43535193 -0.33258154 0.60663818

H -8.29440127 3.03278994 -0.60366291

H -3.81929045 4.29064701 -0.71096652

H -2.91992783 2.82097239 -0.29096857

H -3.56196572 3.86826399 0.98863993

H -4.29350549 0.77641636 0.83672355

**$T_1$ ($^3\pi\pi^*$) min**

C -6.30975876 3.40310125 -0.06493765

C -5.08240594 2.72160665 0.10644923

C -5.12666568 1.28822037 0.48824621

C -7.52826073 1.25242092 -0.06489913

C -3.77720762 3.41169087 -0.06506927

N -6.27828238 0.64333705 0.02776238

N -7.48384035 2.61952101 0.05776247

O -8.54964496 0.60490282 -0.23766875

O -6.42962368 4.64753810 -0.27078577

H -6.32398191 -0.36840528 -0.01332850

H -8.36464261 3.09887410 -0.08999753

H -3.92535356 4.44927693 -0.37063134

H -3.16879898 2.89034947 -0.81528369

H -3.21129748 3.39200267 0.87593980

H -4.22084436 0.69353404 0.43816757

**$CI_{10}$ ($^3\pi\pi^*/S_0$)**

C -6.14770883 3.56055506 -0.32782093

C -4.89051684 2.83002907 -0.20417191

C -5.02102603 1.77551408 0.73540303

C -7.29193181 1.38166000 0.18462081

C -4.67254726 2.19950186 -1.61460695

N -6.02244181 0.90314692 0.65079384

N -7.29070987 2.69568004 -0.20647603

O -8.27269179 0.65738899 0.25148789

O -6.38332692 4.73886904 -0.57844896

H -6.10387983 0.03493489 1.17638479

H -8.20999688 3.09121903 -0.38133397

H -5.54607823 1.66446490 -2.02344207

H -3.82339826 1.50748186 -1.58943303

H -4.41769722 3.01314791 -2.30099889

H -4.37484441 1.70858635 1.61472235